\documentclass[aps,prb,twocolumn,groupedaddress,showpacs,amsmath,amssymb]{revtex4}

\usepackage[dvips]{graphicx}
\usepackage{latexsym}
\usepackage{rsfs}

\begin{document}


\title{Correlated electron systems periodically driven out of equilibrium:\\
Floquet + DMFT formalism}

\author{Naoto Tsuji, Takashi Oka, and Hideo Aoki}
\affiliation{Department of Physics, University of Tokyo, Hongo, 
Tokyo 113-0033, Japan}


\date{\today}

\begin{abstract}
We propose to combine the Floquet formalism for systems in ac fields 
with the dynamical mean-field theory to study 
correlated electron systems periodically driven 
out of equilibrium by external fields such as intense laser light. 
This approach has a virtue that we can nonperturbatively 
include both the correlation effects and 
nonlinear effects due to the driving field, 
which is imperative in analyzing recent experiments for 
photoinduced phase transitions. 
In solving the problem, we exploit a general theorem 
that the Hamiltonian in a Floquet matrix form can be exactly 
diagonalized for single-band noninteracting systems.  
As a demonstration, we have applied the method to the 
Falicov-Kimball model in intense ac fields to calculate the 
spectral function. The result 
shows that photoinduced midgap states emerge from strong ac 
fields, triggering an insulator-metal transition. 
\end{abstract}



\maketitle

\section{Introduction}
\label{intro}

Controlling phases of matter is a central issue in the physics of strongly correlated 
electron systems, where a rich variety of phases 
are realized for various physical degrees of freedom such as spin and charge. 
Since a phase transition dramatically alters macroscopic properties of 
the system, 
it is of great importance to know 
how the phase changes as external parameters (temperature, pressure, 
band filling of the system, etc.) are varied, in equilibrium.

Now, the last decade witnessed that 
a controlling factor, intense laser fields, 
can trigger ``phase transitions'' in correlated 
electron materials.\cite{nasu2004, tokura2006} 
One representative class of materials is perovskite manganites, 
in which an insulator-to-metal transition is induced by photoexcitation.\cite{miyano_ttt1997, fiebig_mtt1998}
Recent experiments also indicate that a 
ferromagnetic spin alignment can emerge in the induced metallic 
phase in manganites.\cite{matsubara_ootot2007} 
Such phenomena are called photoinduced 
phase transitions (PIPTs), where 
irradiation of photons allows 
one to change electronic, magnetic, optical, or structural properties 
of the system. 

The photoinduced phase transition, however, is distinct from 
conventional phase transitions in equilibrium in that 
the photon field drives the system out of equilibrium. 
In other words, the change in phases in nonequilibrium challenges our 
understanding of phase transitions 
which is normally conceived in equilibrium. 
In the PIPT, 
we have to consider, on top of 
the nonlinear electric-field effect, the electron correlation effect. The 
nonlinear effect appears as a threshold behavior, 
i.e., a macroscopic transition only occurs when the intensity of 
the driving field exceeds a certain strength. 
On the other hand, important 
correlation effects such as Mott's metal-insulator transition 
are nonperturbative effects.  Thus, we cannot employ 
the linear-response theory\cite{kubo1957} nor a mean-field treatment 
of the electron-electron interaction. 

Here we propose a theoretical 
approach\cite{tsuji2008a} for photoinduced phenomena, where the 
Floquet-Green function method 
(FGFM)\cite{faisal1989,althorpe_khm1997,brandes_robinson2002,martinez2003} 
is plugged into the dynamical mean-field theory (DMFT).\cite{georges_kkr1996} 
This formulation provides us with a 
promising way to treat both of the nonlinear 
effect of the electric field and the correlation effect 
simultaneously in a nonperturbative manner. 
We then apply the method to study the response of the 
Falicov-Kimball (FK) model,\cite{falicov_kimball1969, freericks_zlatic2003} 
one of the simplest lattice models for correlated 
electrons, to ac electric fields.  
A particular emphasis is put on the technical basis of FGFM. 

During the preparation of the present study, we notice 
that Joura {\it et al.}\cite{joura_fp2008} adopted 
a technique similar to FGFM to solve the Dyson 
equation for a system in dc fields.  
Here we present a general framework of DMFT out of equilibrium 
for {\it arbitrary time-periodic fields} 
in a more transparent viewpoint exploiting the Floquet formalism. 
In this context we should emphasize that FGFM is not just a numerical 
technique but offers a fruitful physical picture for nonequilibrium systems 
as revealed here.


The paper is organized as follows: 
in Sec.~\ref{flq_theorem}, we review the Floquet theorem and the Floquet 
matrix method, on which our theoretical description is based. 
The rest of the paper is devoted to our original formulation. 
In Sec.~\ref{flq_rep_g} we define the Floquet matrix form of 
Green's function, which is our starting point of FGFM. 
Then in Sec.~\ref{noninteracting}, we derive a general expression for 
Green's function and its inverse for noninteracting electrons in a Floquet matrix 
form. We calculate noninteracting Green's functions for several cases 
to discuss their physical implications. 
In Sec.~\ref{diagonalization}, we derive a general theorem that 
identifies the eigenvalues and the eigenvectors of a Floquet matrix form 
of the Hamiltonian for noninteracting electrons. 
While Secs.~\ref{noninteracting} and \ref{diagonalization} address 
noninteracting cases, we move on 
to correlated electron systems 
by incorporating FGFM in DMFT in Sec.~\ref{dmft}. 
We formulate the obtained Green's function in a gauge-invariant manner 
in Sec.~\ref{gauge}. 
We then apply our method to the FK model in Sec.~\ref{fk}, 
and calculate the spectral functions for dc and ac fields, 
where we discuss how the Mott-insulating state is transformed 
into the metallic states in the external fields. 
We summarize the paper and give future problems in Sec.~\ref{conclusion}.


\section{Floquet theorem and Floquet matrix}
\label{flq_theorem}

An external field drives an 
electron having an energy $\varepsilon$ into another state with a different energy, 
where there are many scattering channels in nonequilibrium. 
However, if the driven system is periodic in time with a frequency $\Omega$, the allowed channels 
are limited to the processes such that $\varepsilon \to \varepsilon + n\Omega$, 
where $n$ is an integer. 
This greatly reduces the channels' degrees of freedom to be dealt with. 
One can take advantage of such a simplification through the Floquet 
matrix method,\cite{shirley1965, sambe1973, dittrich_hiksz1998} 
of which we give an overview in this section. 

The method has been used as an effective 
approach toward photoexcited systems. 
The concept of the Floquet matrix originates from the 
Floquet theorem \cite{floquet1883} for a periodically driven system, an analog of the Bloch theorem
applied to a spatially periodic system. 
Floquet theorem is a general theorem for differential equations of a 
form $dx(t)/dt=C(t)x(t)$ with $C$ periodic in $t$, which include 
equations of motion for systems subject to external driving forces 
that periodically oscillate in time. 
One representative example is the Mathieu equation which describes 
a parametric resonance phenomenon. 
Here we restrict ourselves to a quantum system 
whose dynamics is determined by the time-dependent Schr\"{o}dinger equation, 
\begin{equation}
  i\frac{d}{dt} \Psi(t)
	  =
		  H(t) \Psi(t),
  \label{schrodinger}
\end{equation}
where $\Psi(t)$ is a state vector of the system, 
and $H(t)$ is the time-dependent Hamiltonian, which is assumed 
to be periodic in $t$, 
\begin{equation}
  H(t+\tau)
	  =
		  H(t),
  \label{periodicity}
\end{equation}
with a period $\tau$. 
The Floquet theorem states that 
there exists a solution of Eq.~(\ref{schrodinger}) which 
is an eigenstate of the time translation operation $t \to t + \tau$, 
implying 
\begin{equation}
  \Psi_\alpha(t)
	  =
		  e^{-i\varepsilon_\alpha t} u_\alpha(t)
	\label{floquet_state}
\end{equation}
with $e^{-i\varepsilon_\alpha \tau}$ as an eigenvalue of the time translation, 
$\alpha$ as a set of quantum numbers, and $u_\alpha(t)=u_\alpha(t+\tau)$ 
as a periodic function of $t$. Hence we can Fourier expand $u_\alpha(t)$ as 
$u_\alpha(t) = \sum_n e^{-in\Omega t}\, u_\alpha^n$
with the frequency $\Omega = 2\pi/\tau$, where $u_\alpha^n$ is called 
the $n$th Floquet mode of Floquet state (\ref{floquet_state}). 
We can then Fourier transform Eq.~(\ref{schrodinger}) to have
\begin{equation}
  \sum_n H_{mn} u_\alpha^n
	  =
		  (\varepsilon_\alpha+m\Omega) u_\alpha^m,
	\label{floquet}
\end{equation}
where 
\begin{equation}
  H_{mn}
	  \equiv 
		  \frac{1}{\tau}\int_{-\tau/2}^{\tau/2} dt \;
			e^{i(m-n)\Omega t} H(t)
	\label{floquet_hamiltonian}
\end{equation}
is the Floquet matrix form of the Hamiltonian. 
The factor $\varepsilon_\alpha+m\Omega$ appearing on the 
right-hand side (rhs) of 
Eq.~(\ref{floquet}) is called {\it quasienergy}, 
which forms a ladder of energies with a spacing $\Omega$. 
Since the Hamiltonian is time dependent, the energy is not conserved in general. 
However, Eq.~(\ref{floquet}) shows that the energy is conserved up to an 
integer multiple of $\Omega$, corresponding to the absorption or emission 
of the photon with the energy $\Omega$. 
Each element in the Floquet matrix $H_{mn}$ corresponds to the probability amplitude 
of the transition from the $m$th Floquet mode to the $n$th one, so that 
off-diagonal components represent excitations driven by the external 
field while the diagonal ones the probability to remain in the same mode. 

The consequence of the Floquet theorem is 
remarkable: Eq.~(\ref{floquet}) resembles the static Schr\"{o}dinger equation in 
equilibrium except for the presence of the Floquet mode index $n$, 
which means that we have no longer to solve the 
time-dependent Schr\"{o}dinger Eq.~(\ref{schrodinger}), 
in favor of the {\it time-independent} Eq.~(\ref{floquet}). 
This is the great advantage of the Floquet matrix method, which also plays 
a crucial role in Green's-function approach.

\section{Floquet representation of Green's function}
\label{flq_rep_g}

Besides the Floquet analysis of the Hamiltonian, 
we can alternatively describe 
nonequilibrium states in Green's-function approach 
based on the Keldysh formalism.\cite{schwinger1961, keldysh1964} 
The approach of the Floquet matrix proves its own worth when it is used within 
Green's-function formalism. 
The idea of FGFM 
was first introduced by Faisal,\cite{faisal1989} 
followed by several groups.\cite{althorpe_khm1997, brandes_robinson2002, martinez2003} 
In this section we give another way to define a Floquet matrix form of 
Green's function, which we shall use in this paper. 


A Green's function has 
two independent arguments of time, $t$ and $t'$, as denoted by $G(t, t')$. 
We define variables $t_{\rm rel}\equiv t-t'$ and $t_{\rm av}\equiv (t+t')/2$. 
In equilibrium the 
system is invariant against continuous time translation, so that 
Green's functions depend on $t$ and $t'$ only through $t_{\rm rel}$, which 
enables us to Fourier transform them into functions of $\omega$. 
However, when the system is driven out of equilibrium, they generally depend on 
both $t_{\rm rel}$ and $t_{\rm av}$. 
Since the periodic system that we consider in this paper has the discrete time 
translation invariance [Eq.~(\ref{periodicity})], 
it is guaranteed that Green's function is also invariant against $t_{\rm av} 
\to t_{\rm av} + \tau$. For an arbitrary function $G(t, t')$ (not limited 
to Green's function) that satisfies the periodicity condition, 
$G(t+\tau, t'+\tau) = G(t, t')$, we can define the Wigner transformation
of $G$ as 
\begin{equation}
  G_n(\omega)
	  =
		  \int_{-\infty}^{\infty} dt_{\rm rel} \;
			\frac{1}{\tau}\int_{-\tau/2}^{\tau/2} dt_{\rm av} \;
			e^{i\omega t_{\rm rel}+in\Omega t_{\rm av}}
			G(t, t').
	\label{wigner_rep}
\end{equation}
We call $G_n(\omega)$ the {\it Wigner representation} of the function $G$. 
Using the Wigner representation, we define the Floquet matrix form of $G$ as 
\begin{equation}
  G_{mn}(\omega)
	  \equiv 
		  G_{m-n}\!\left( \omega+\frac{m+n}{2}\Omega \right), 
	\label{floquet_rep}
\end{equation}
and call $G_{mn}(\omega)$ the {\it Floquet representation}. Hereafter a function 
with one index $n$ should be understood as a Wigner representation, while 
two indices $m$ and $n$ mean a Floquet representation. 
In the Floquet representation, we use the reduced zone scheme, i.e., 
the range of $\omega$ is restricted 
to the ``Brillouin zone'' on the frequency axis: $-\Omega/2 < \omega \leq \Omega/2$. 
We can readily check that definition (\ref{floquet_rep}) is 
equivalent to the one given by Refs.~\onlinecite{faisal1989, althorpe_khm1997, brandes_robinson2002, martinez2003}.
The Floquet representation is used during calculations, while the Wigner representation is used 
when we interpret the result, 
since the Wigner representation has a clear physical interpretation that 
$G_n$ is the $n$th oscillating mode in $t_{\rm av}$ of $G(t,t')$. 

Actually, the Floquet representation $G_{mn}(\omega)$ 
has a one-to-one correspondence with 
the Wigner representation $G_{\ell}(\omega ')$. $G_{\ell}(\omega ')
\to G_{mn}(\omega)$: the integers $m$ and $n$ should obey the conditions 
\begin{align}
	  & m-n = \ell,
		\label{wig_flq1}
	\\
    -\frac{\Omega}{2} <\; & \omega '- \frac{m+n}{2}\Omega \leq \frac{\Omega}{2},
		\label{wig_flq2}
\end{align}
for $\ell$ and $\omega '$. There are two consecutive integers $k$ and 
$k+1$ which can be equal to $m+n$ satisfying Eq.~(\ref{wig_flq2}). 
Either $k$ or $k+1$ is congruent to $\ell$ modulo 2. Thus $m+n$ is 
uniquely determined via Eq.~(\ref{wig_flq2}) and the condition 
$m+n \equiv m-n \equiv \ell \; ({\rm mod}\; 2)$. Together with Eq.~(\ref{wig_flq1}), 
$m$ and $n$ are uniquely determined. For such $m$ and $n$, $\omega$ 
is given by $\omega '-(m+n)\Omega/2$. 
$G_{mn}(\omega) \to G_{\ell}(\omega ')$: $\ell$ and $\omega '$ are 
uniquely determined by $\ell = m-n$ and $\omega '= \omega+(m+n)\Omega/2$. 

We can immediately realize 
the advantage of the Floquet representation 
in multiplications of two 
Floquet-represented functions. As shown in Appendix \ref{multiplication},
the mapping from $G(t, t')$ to $G_{mn}(\omega)$ preserves a 
multiplication structure, 
\begin{align*}
    &\int dt'' A(t,t'') B(t'', t') 
		  = 
			  C(t, t'), 
	\\
	  &\Leftrightarrow 
	  \sum_\ell A_{m\ell}(\omega) B_{\ell n}(\omega)
		  =
			  C_{mn}(\omega).
\end{align*}
As an example, the Floquet representation of the Dyson equation 
is simplified into 
\begin{align}
  (G_{\boldsymbol k})_{mn}(\omega)
	  &=
		  (G_{\boldsymbol k}^0)_{mn}(\omega) 
      + \sum_{m'n'}
			(G_{\boldsymbol k}^0)_{mm'}(\omega) (\Sigma_{\boldsymbol k})_{m'n'}(\omega)
	\nonumber
	\\
	  &\quad \times
			(G_{\boldsymbol k})_{n'n}(\omega),
	\label{dyson}
\end{align}
where $G$ and $G^0$ are, respectively, 
the full and the noninteracting Green's functions 
and $\Sigma$ is the self-energy. 
Note that each function has the additional $2\times 2$ matrix structure, 
$G=\begin{pmatrix} G^R & G^K \\ 0 & G^A\end{pmatrix}$
in the Keldysh space (with the three linearly independent components: 
the retarded, the advanced, and the Keldysh one). 
Thanks to the usual multiplication rule of 
the linear algebra, one can solve the Dyson Eq.~(\ref{dyson}) as 
$G_{\boldsymbol k}(\omega) = [{G_{\boldsymbol k}^0}^{-1}(\omega)-\Sigma_{\boldsymbol k}(\omega)]^{-1}$.
In addition, a typical size of a Floquet matrix that is needed to be taken in 
numerical calculations is usually small because 
sufficiently high-order processes should tend to be irrelevant when 
the driving field is not so large, which also supports the usefulness of 
FGFM. 

\section{Noninteracting electrons}
\label{noninteracting}

Having defined the Floquet representation of Green's function in 
Eq.~(\ref{floquet_rep}), we then compute the Floquet-represented 
Green's function for noninteracting electrons. 
Although FGFM has been used by several authors to study 
noninteracting electrons driven out of equilibrium, 
there are still further developments yet to be explored. 
This has motivated us to present an exact and unified description of 
FGFM that can be applied to general noninteracting single-band systems 
in this section. 

In Sec.~\ref{general_lattice_field} we provide a general expression of the 
Floquet representation of Green's function for {\it any} single-band model. Then in 
Sec.~\ref{inverse_green} we derive the inverse of Green's 
function, which will be used to build DMFT in the Floquet matrix 
form in Sec.~\ref{dmft}. After that, we show several examples of Green's function for the 
hypercubic (Sec.~\ref{hypercubic}) and other lattices 
(Sec.~\ref{other_lattice}). Throughout the paper we restrict our discussion 
to a single-band model, and omit spin degrees of freedom for simplicity. 

\subsection{General lattices and fields}
\label{general_lattice_field}

Let $\epsilon_{\boldsymbol k}$ be a band dispersion of the system. We make 
the system subject to a homogeneous time-dependent electric field 
periodic in $t$. 
Here we choose the temporal gauge or the 
Hamiltonian gauge in which the scalar potential $\phi = 0$. 
Replacing the momentum $\boldsymbol k$ 
with ${\boldsymbol k}-e{\boldsymbol A}(t)$ [${\boldsymbol A}(t)$: a vector potential]
in $\epsilon_{\boldsymbol k}$ 
gives the noninteracting Hamiltonian, 
\begin{equation}
  H(t)
	  =
		  \sum_{\boldsymbol k} \epsilon_{{\boldsymbol k}-e{\boldsymbol A}(t)} 
			c_{\boldsymbol k}^\dagger c_{\boldsymbol k},
\end{equation}
where $c_{\boldsymbol k}^\dagger$ and $c_{\boldsymbol k}$ are the creation and the 
annihilation operators of the electrons, respectively, and we treat the 
electric field as a classical one. The retarded Green's function for 
noninteracting electrons reads 
\begin{align}
  G_{\boldsymbol k}^{R0}(t,t')
    &=
      -i\theta(t-t')
      \langle[
        c_{\boldsymbol k}(t),\;c_{\boldsymbol k}^\dagger(t')
      ]_+\rangle_0
  \nonumber
  \\
    &=
      -i\theta(t-t')
      \exp\left(
        -i\int_{t'}^t dt'' \;
				[ \epsilon_{{\boldsymbol k}-e{\boldsymbol A}(t'' )}-\mu ]
      \right),
  \label{retarded1}
\end{align}
where $\theta(t)$ represents the step function, 
$[\;, \;]_+$ is the anticommutation relation, $\langle \cdots \rangle_0$ is
the statistical average with respect to the initial density matrix 
$\rho_0 = e^{-\beta H({\boldsymbol A}=0)}$ (where the system is assumed to be in 
equilibrium with the temperature $\beta^{-1}$ at $t=-\infty$), 
and $\mu$ is the chemical potential of the system. We can transform 
Eq.~(\ref{retarded1}) into the Wigner representation via Eq.~(\ref{wigner_rep}), and then into the 
Floquet representation through Eq.~(\ref{floquet_rep}). The details of 
the calculation are described in Appendix \ref{flq_g}. 
The final result is 
\begin{align}
  (G_{\boldsymbol k}^{R0})_{mn}(\omega)
	  =&
		  \sum_\ell \frac{1}{\omega+\ell\Omega+\mu-(\epsilon_{\boldsymbol k})_0+i\eta}
		\nonumber
  \\
    &\times
			\int_{-\pi}^{\pi} \frac{dx}{2\pi} \int_{-\pi}^{\pi} \frac{dy}{2\pi} \;
			e^{i(m-\ell)x+i(\ell-n)y}
	\nonumber
	\\
		&\times\exp\left(
			  -\frac{i}{\Omega}\int_{y}^{x} dz \;
				[
				  \epsilon_{{\boldsymbol k}-e{\boldsymbol A}(z/\Omega)}-(\epsilon_{\boldsymbol k})_0
				]
		  \right),
		\nonumber
  \\
	\label{floquet1}
\end{align}
where $\eta$ is a positive infinitesimal, and
$(\epsilon_{\boldsymbol k})_0$ is the time-averaged dispersion, 
which coincides with the zeroth Floquet mode of $\epsilon_{\boldsymbol k}$, 
\begin{equation}
  (\epsilon_{\boldsymbol k})_{m-n}
	  =
		  \int_{-\pi}^{\pi} \frac{dz}{2\pi} \;
			e^{i(m-n)z} \epsilon_{{\boldsymbol k}-e{\boldsymbol A}(z/\Omega)}.
	\label{e_flq}
\end{equation}
Equation (\ref{floquet1}) is the general Floquet representation of Green's function for 
the noninteracting system driven by a periodic field. What is notable about 
expression (\ref{floquet1}) is that it can be decomposed into 
well-behaved matrices. Let us define two Floquet matrices, 
\begin{align}
  (\Lambda_{\boldsymbol k})_{mn}
    &=
      \int_{-\pi}^\pi \frac{dx}{2\pi}\;e^{i(m-n)x}
	\nonumber
	\\
    &\times\exp\left(
        -\frac{i}{\Omega}\int_0^x dz\;
        [
				  \epsilon_{{\boldsymbol k}-e{\boldsymbol A}(z/\Omega)}-(\epsilon_{\boldsymbol k})_0
				]
      \right) ,
	\label{lambda}
\end{align}
and 
\begin{equation}
  (Q_{\boldsymbol k})_{mn}(\omega)
    =
      \frac{1}{\omega+n\Omega+\mu-(\epsilon_{\boldsymbol k})_0+i\eta}\;\delta_{mn},
  \label{q}
\end{equation}
where $\Lambda_{\boldsymbol k}$ is unitary as shown in Appendix \ref{unitarity}, 
and $Q_{\boldsymbol k}(\omega)$ is a diagonal matrix. The physical meaning of 
these matrices will be given later in Sec.~\ref{diagonalization}. 
Here let us just note a strikingly simple decomposition, 
\begin{equation}
  G_{\boldsymbol k}^{R0}(\omega)
	  =
		  \Lambda_{\boldsymbol k} \cdot Q_{\boldsymbol k}(\omega) \cdot \Lambda_{\boldsymbol k}^\dagger.
	\label{lambda_q_lambda}
\end{equation}
where we denote a multiplication of a Floquet matrix by `` $\cdot$,'' and 
omit Floquet indices in Eq.~(\ref{lambda_q_lambda}). 
The decomposition [Eq.~(\ref{lambda_q_lambda})] is essentially used to derive the inverse of Green's 
function in Sec.~\ref{inverse_green}. 

The Floquet representation of the 
advanced Green's function is equal to the Hermitian adjoint of the 
retarded one: 
$(G_{\boldsymbol k}^{A0})_{mn}(\omega) = ({G_{\boldsymbol k}^{R0}}^\dagger)_{mn}(\omega)
= (G_{\boldsymbol k}^{R0})_{nm}^\ast(\omega)$. Using Eq. 
(\ref{lambda_q_lambda}), we have $G_{\boldsymbol k}^{A0}(\omega) = 
\Lambda_{\boldsymbol k} \cdot Q_{\boldsymbol k}^\dagger(\omega) \cdot 
\Lambda_{\boldsymbol k}^\dagger$. 

\subsection{Inverse of Green's function}
\label{inverse_green}

When one solves a Dyson equation such as Eq.~(\ref{dyson}) to include effects of interaction, 
the noninteracting part appears as an inverse, ${G_{\boldsymbol k}^{R0}}^{-1}(\omega)$, 
rather than $G_{\boldsymbol k}^{R0}(\omega)$ itself. 
Using relation (\ref{lambda_q_lambda})
and the unitarity of $\Lambda_{\boldsymbol k}$,
we can analytically invert Green's function as 
${G_{\boldsymbol k}^{R0}}^{-1}(\omega) = \Lambda_{\boldsymbol k} \cdot 
Q_{\boldsymbol k}^{-1}(\omega) \cdot \Lambda_{\boldsymbol k}^\dagger$, which can 
be evaluated exactly as presented in Appendix \ref{inverse_g}. 
The derived expression for the inverse of Green's function is 
\begin{equation}
  ({G_{\boldsymbol k}^{R0}}^{-1})_{mn}(\omega)
	  =
		  (\omega+n\Omega+\mu+i\eta) \delta_{mn}
			- (\epsilon_{\boldsymbol k})_{m-n}.
	\label{inv_g}
\end{equation}
One can also prove Eq.~(\ref{inv_g}) from the Dyson equation for 
$G_{\boldsymbol k}^{R0}(t, t')$ in a straightforward manner. 
Relation (\ref{inv_g}) means that 
Green's function is the kernel of Eq.~(\ref{floquet}), or that 
{\it the Floquet representation of Green's function is equivalent 
to the inverse of the Floquet matrix form of the quasienergy minus the Hamiltonian.} 
One can use Eq.~(\ref{inv_g}) for {\it any} single band Hamiltonian 
with a homogeneous electric field periodic in time. 
In Secs.~\ref{hypercubic} and \ref{other_lattice}, we present examples of the calculations that 
utilize relation (\ref{inv_g}). 

\subsection{Hypercubic lattice}
\label{hypercubic}

As a first example, let us consider a simple cubic lattice in $d$ 
dimensions, whose energy dispersion is given by 
\begin{equation}
  \epsilon_{\boldsymbol k}^{\rm sc}
	  =
		  -2t \sum_{i=1}^d \cos k_i,
  \label{e_sc}
\end{equation}
where $t$ is the hopping and we set the lattice constant $a=1$. 
For simplicity we assume that the 
vector potential ${\boldsymbol A}(t)$ is parallel to $(1,1,\dots,1)$ with
each component $A_i(t) = A(t)$. Substituting $k_i$ with 
$k_i-eA(t)$ in Eq.~(\ref{e_sc}), we have 
\begin{equation}
  \epsilon_{{\boldsymbol k}-e{\boldsymbol A}(t)}^{\rm sc}
	  =
		  \epsilon_{\boldsymbol k}^{\rm sc} \cos eA(t)
			+ \bar{\epsilon}_{\boldsymbol k}^{\rm sc} \sin eA(t),
\end{equation}
where we have defined 
\begin{equation}
  \bar{\epsilon}_{\boldsymbol k}^{\rm sc}
	  =
		  -2t \sum_{i=1}^d \sin k_i,
	\label{ebar_sc}
\end{equation}
after Turkowski and Freericks.\cite{turkowski_freericks2005} Note that 
$\bar{\epsilon}_{\boldsymbol k}^{\rm sc}$ has the odd time-reversal symmetry. 
Every equation including $\epsilon_{\boldsymbol k}^{\rm sc}$ and 
$\bar{\epsilon}_{\boldsymbol k}^{\rm sc}$ must be consistent against 
the time-reversal operation. For instance, one usually finds the factor 
$\epsilon_{\boldsymbol k}^{\rm sc}+i\bar{\epsilon}_{\boldsymbol k}^{\rm sc}$, 
which is time-reversal even, 
since the imaginary unit $i$ is time-reversal odd. 

An integral over ${\boldsymbol k}$ is performed through 
\begin{equation}
  \rho(\epsilon, \bar{\epsilon})
    =
      \sum_{\boldsymbol k}
      \delta(\epsilon-\epsilon_{\boldsymbol k}^{\rm sc})
      \delta(\bar{\epsilon}-\bar{\epsilon}_{\boldsymbol k}^{\rm sc}),
	\label{jdos_def}
\end{equation}
which is called the joint density of states (JDOS).\cite{turkowski_freericks2005} 
This contrasts with the equilibrium cases in which an integrand depends on ${\boldsymbol k}$ only through 
$\epsilon_{\boldsymbol k}^{\rm sc}$, so that we can replace the integral 
variable ${\boldsymbol k}$ with $\epsilon_{\boldsymbol k}^{\rm sc}$ accompanied by 
the usual density of states $\rho(\epsilon)$. The analytic expression 
for JDOS in arbitrary $d$ 
dimensions is summarized in Appendix \ref{jdos}. In particular, in 
infinite dimensions the JDOS becomes a Gaussian function 
[Eq.~(\ref{gaussian})].\cite{turkowski_freericks2005} 
In the following, we consider two kinds of electric fields, a dc field 
(Sec.~\ref{hc_dc}), and an ac field (Sec.~\ref{hc_ac}). 

\subsubsection{Hypercubic lattice in a dc field}
\label{hc_dc}

A homogeneous dc field is given by the vector potential proportional to 
time, 
\begin{equation}
  eA(t)a
    =
      \Omega t,
	\label{dc_vector_p}
\end{equation}
where 
\begin{equation}
  \Omega
	  =
		  -eEa 
	\label{dc_omega}
\end{equation}
is the Bloch frequency. The system with the field 
[Eq.~(\ref{dc_vector_p})] fulfills the periodicity condition [Eq.~(\ref{periodicity})]
with the period $2\pi/\Omega$ because of the periodic potential of the 
lattice. With Eq.~(\ref{floquet1}), Green's function 
has the following Floquet representation, 
\begin{align}
  (G_{\boldsymbol k}^{R0})_{mn}(\omega)
	  &=
		  e^{i(m-n)\theta_{\boldsymbol k}}
			\sum_\ell \frac{1}{\omega+\ell\Omega+\mu+i\eta}\;
	\nonumber
	\\
	  &\quad\times
			J_{m-\ell}\!\left(\frac{\zeta_{\boldsymbol k}}{\Omega}\right)
			J_{\ell-n}\!\left(\frac{\zeta_{\boldsymbol k}}{\Omega}\right),
	\label{sc_dc_flq}
\end{align}
where $J_n(z)$ is the $n$th-order Bessel function, and 
\begin{align}
  &\zeta_{\boldsymbol k}
    =
      \sqrt{(\epsilon_{\boldsymbol k}^{\rm sc})^2+(\bar{\epsilon}_{\boldsymbol k}^{\rm sc})^2},
	\label{zeta_theta1}
	\\
  &\tan\theta_{\boldsymbol k}
    =
      \bar{\epsilon}_{\boldsymbol k}^{\rm sc}/\epsilon_{\boldsymbol k}^{\rm sc}.
	\label{zeta_theta2}
\end{align}
To see how Green's function (\ref{sc_dc_flq}) behaves, we calculate the local spectral function 
$A_n(\omega)=-\frac{1}{\pi}{\rm Im}\sum_{\boldsymbol k} (G_{\boldsymbol k}^{R0})_n(\omega)$
in infinite dimensions. 
First, we transform Eq.~(\ref{sc_dc_flq}) into the Wigner representation, 
and then integrate it over ${\boldsymbol k}$ with the JDOS [Eq.~(\ref{gaussian})]. 
After taking the imaginary part, we arrive at 
\begin{equation}
    A_n(\omega)
		  =
			  \delta_{n, 0} \sum_\ell a_\ell\; \delta(\omega+\ell\Omega+\mu). 
		\label{spec_dc}
\end{equation}
Here the coefficients in front of the delta functions are 
$a_\ell = e^{-1/2\Omega^2}I_{\ell}(1/2\Omega^2)$, 
where $I_\ell(z)$ is the modified Bessel function of the first kind. 
Note that the coefficients satisfy the normalization condition, 
$\sum_\ell a_\ell = 1$.

\begin{figure}[t]
  \begin{center}
	  \includegraphics[width=8cm]{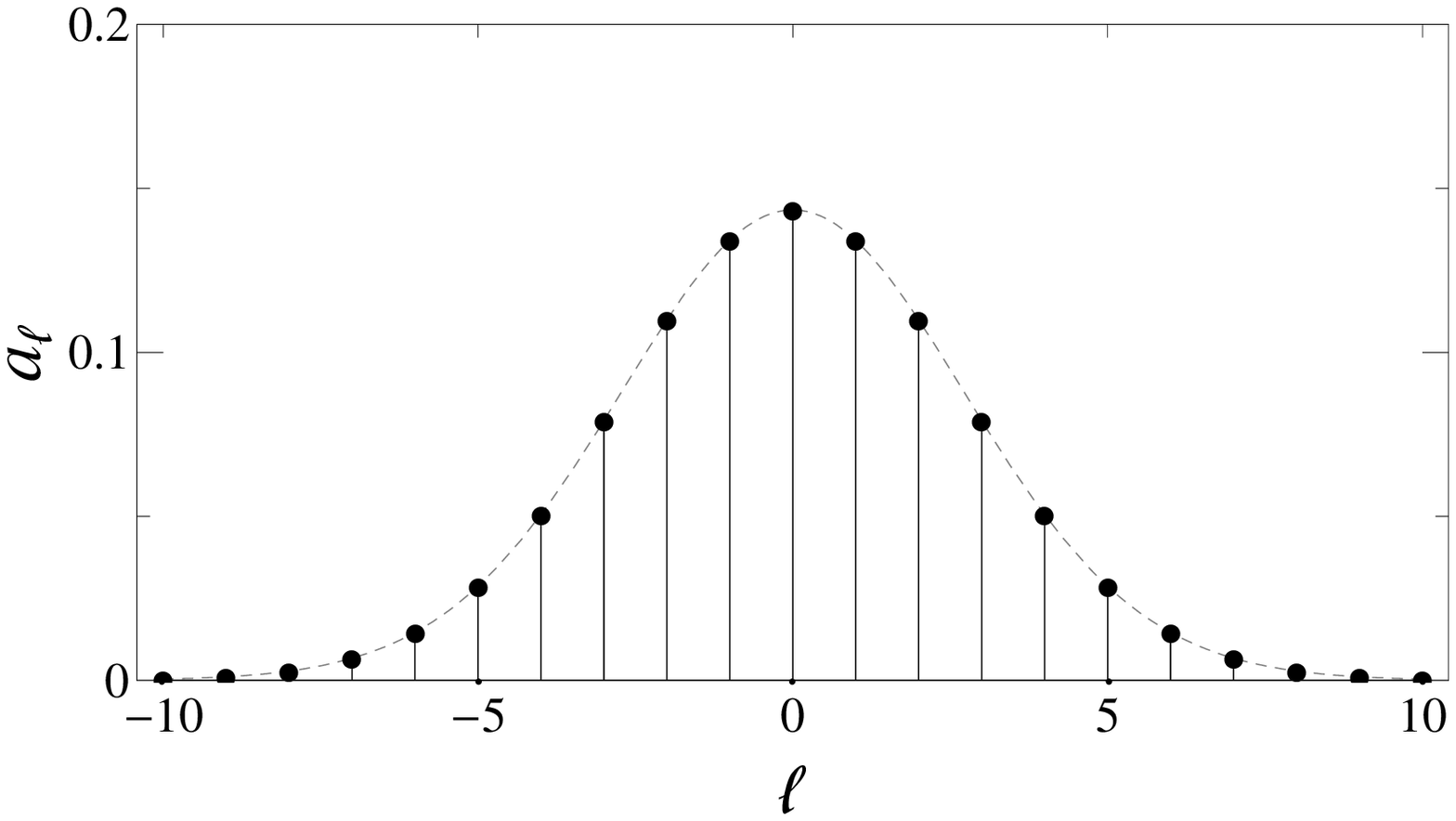}
		\caption{The coefficients $a_\ell$ of the delta 
    functions in the local spectral function of the noninteracting 
	  electrons on the hypercubic lattice 
    with the dc field $\Omega = 0.25$ are plotted by the circles. The 
	  delta functions with the spacing $\Omega$ are schematically shown by 
	  the solid lines. The broken line is a guide for the eyes. }
		\label{wannier_stark}
	\end{center}
\end{figure}
Equation (\ref{spec_dc}) shows that the local spectral function has only 
the $n=0$ component, indicating that the spectral 
function evolves into a time-independent function [Eq.~(\ref{spec_dc})] 
for a sufficiently long time elapsed after the dc field began to drive 
the system.\cite{turkowski_freericks2005} Later in Sec.~\ref{gauge_dc}, 
we shall show that the components with $n\neq 0$ vanish due to a symmetry in 
the system. 

The spectral function on the hypercubic lattice, displayed in Fig.~\ref{wannier_stark},
consists of a set of delta functions with a spacing $\Omega$, namely, the 
Wannier-Stark ladder. 
The width of each peak (approximately effective hopping) is infinitesimal 
due to the Bloch oscillation, where 
an electron is not free to run in a lattice, but only oscillates with the 
frequency $\Omega$. 

The inverse of the Floquet-represented
Green's function in this case is given via relation 
(\ref{inv_g}) by 
\begin{widetext}
\begin{align}
  ({G_{\boldsymbol k}^{R0}}^{-1})_{mn}(\omega)
    &=
      (\omega+n\Omega+\mu+i\eta)\delta_{mn}
      -\frac{1}{2}
      [(\epsilon_{\boldsymbol k}^{\rm sc}+i\bar{\epsilon}_{\boldsymbol k}^{\rm sc})\delta_{m-n,1} 
      + (\epsilon_{\boldsymbol k}^{\rm sc}-i\bar{\epsilon}_{\boldsymbol k}^{\rm sc})\delta_{m-n,-1}]
	\nonumber
	\\
    &=
      (\omega+n\Omega+\mu+i\eta)\delta_{mn}
			-\frac{1}{2}e^{i(m-n)\theta_{\boldsymbol k}}\zeta_{\boldsymbol k}
      (\delta_{m-n,1}+\delta_{m-n,-1}).
	\label{sc_dc1}
\end{align}
The Hamiltonian part of Eq.~(\ref{sc_dc1}) is explicitly written down in 
a tridiagonal matrix form as 
\begin{equation}
  \frac{1}{2}
    \begin{pmatrix}
      \ddots & \ddots & & & & & & 
      \\
      \ddots & 0 & \zeta_{\boldsymbol k}e^{i\theta_{\boldsymbol k}} & & & 0 & &
      \\
       & \zeta_{\boldsymbol k}e^{-i\theta_{\boldsymbol k}} & 0 & 
         \zeta_{\boldsymbol k}e^{ i\theta_{\boldsymbol k}} & & & &
      \\
       & & \zeta_{\boldsymbol k}e^{-i\theta_{\boldsymbol k}} & 0 & 
           \zeta_{\boldsymbol k}e^{ i\theta_{\boldsymbol k}} & & &
      \\
       & & & \zeta_{\boldsymbol k}e^{-i\theta_{\boldsymbol k}} & 0 & 
             \zeta_{\boldsymbol k}e^{ i\theta_{\boldsymbol k}} & &
      \\
       & 0 & & & \zeta_{\boldsymbol k}e^{-i\theta_{\boldsymbol k}} & 0 & 
               \ddots &
      \\
       & & & & & \ddots & \ddots
		\end{pmatrix}.
	\label{dc_matrix}
\end{equation}
\end{widetext}
As remarked in Sec.~\ref{flq_theorem}, each component of the Floquet matrix 
[Eq.~(\ref{dc_matrix})] represents a probability amplitude of a transition 
from one Floquet mode to another. Note that 
for the case of the dc field, Hamiltonian (\ref{dc_matrix})
has no diagonal components, which means that the electrons cannot stay 
stationary but are always excited by the field. We also note that the 
off diagonal components do not depend on $\Omega$ which is proportional to the 
strength of the field. The dependence of $\Omega$ is only taken into 
account through the quasienergy part of Eq.~(\ref{sc_dc1}). 

\subsubsection{Hypercubic lattice in an ac field}
\label{hc_ac}

\begin{figure}[t]
  \begin{center}
	  \includegraphics[width=8cm]{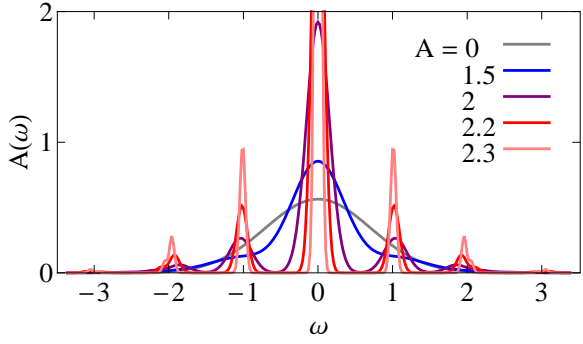}
    \caption{(Color online) The local spectral functions of the noninteracting 
	  electrons on the hypercubic lattice 
    coupled to the ac field with $\Omega=1$ and $A=0, 
	  1.5, 2, 2.2$, and $2.3$. }
    \label{ac_spec_free_fig}
		\end{center}
\end{figure}

Let us move on to the case of the ac field on the hypercubic lattice. 
The vector potential is defined by 
\begin{equation}
  eA(t)a
    =
      A\sin\Omega t, 
\end{equation}
where $\Omega$ is the frequency of the ac field, and 
\begin{equation}
  A
	  =
		  -\frac{eEa}{\Omega}
	\label{ac_a}
\end{equation}is its 
amplitude divided by the frequency. Although we use the symbol $\Omega$, 
this should not be confused with $\Omega$ (the Bloch frequency) 
introduced in Eq.~(\ref{dc_omega}) for the dc field. Following 
Eq.~(\ref{floquet1}), we derive the Floquet representation of Green's 
function as 
\begin{align}
  (G_{\boldsymbol k}^{R0})_{mn}(\omega)
    &=
      \sum_{\ell}
      \frac{1}{\omega+\ell\Omega+\mu-\epsilon_{\boldsymbol k}^{\rm sc}J_0(A)+i\eta}
	\nonumber
	\\
	  &\times
      \int_{-\pi}^{\pi}\frac{dx}{2\pi}\;
      \int_{-\pi}^{\pi}\frac{dy}{2\pi}\;
      e^{i(m-\ell)x+i(\ell-n)y}
  \nonumber
  \\
    &\times
      \exp\bigg(
        -\frac{i}{\Omega}\int_{y}^{x}
        dz\; 
        \{\epsilon_{\boldsymbol k}^{\rm sc}[\cos(A\sin z)-J_0(A)]
	\nonumber
	\\
	  &
        +\bar{\epsilon}_{\boldsymbol k}^{\rm sc}\sin(A\sin z)\}
      \bigg) .
	\label{sc_ac_flq}
\end{align}
Taking the imaginary part of Eq.~(\ref{sc_ac_flq}) and integrating over 
${\boldsymbol k}$ with the JDOS [Eq.~(\ref{gaussian})], we obtain the local spectral 
function. We depict it for several $A$ and $\Omega=1$
in Fig.~\ref{ac_spec_free_fig}. One can see that the spectral function 
has narrow peaks at $\omega=n\Omega$ ($n=0, \pm 1, \pm 2, \dots$) 
known as the {\it dynamical Wannier-Stark ladder}. 
The peaks at $\omega=\pm\Omega$ correspond to one-photon 
absorption or emission, and the peaks at $\omega=\pm 2\Omega$ 
correspond to two-photon 
one, etc. The width of each peak, or the effective hopping, is renormalized 
by the zeroth-order Bessel function $J_0(A)$ as seen in Eq.~(\ref{sc_ac_flq}), and even 
vanishes making the electrons completely localized when $J_0(A)=0$. 
In Fig.~\ref{ac_spec_free_fig} one finds that the widths of the peaks 
shrink as $A$ approaches the first zero ($z=2.40483\cdots$) of 
$J_0(A)$ until finally the peaks become the delta functions. 
This scaling has been known as {\it dynamical localization} since 
the proposal by Dunlap and Kenkre.\cite{dunlap_kenkre1986} 

The inverse of Green's function in the ac field can be calculated via 
Eq.~(\ref{inv_g}). The result is 
\begin{widetext}
\begin{equation}
  ({G_{\boldsymbol k}^{R0}}^{-1})_{mn}(\omega)
    =
      (\omega+n\Omega+\mu+i\eta)\delta_{mn}-
        \begin{cases}
          \epsilon_{\boldsymbol k}^{\rm sc}J_{m-n}(A) & 
          m-n{\rm : even}
          \\
          i\,\bar{\epsilon}_{\boldsymbol k}^{\rm sc}J_{m-n}(A) & 
					m-n{\rm : odd}
        \end{cases}.
	\label{sc_ac}
\end{equation}
The explicit Floquet matrix form of the Hamiltonian part of Eq.~(\ref{sc_ac}) reads
\begin{equation}
    \begin{pmatrix}
      \ddots & & &\vdots & & & 
      \\
      \vspace{.2cm}
       & \epsilon_{\boldsymbol k}J_0(A) & i\bar{\epsilon}_{\boldsymbol k}J_1(A) & 
         \epsilon_{\boldsymbol k}J_2(A) & i\bar{\epsilon}_{\boldsymbol k}J_3(A) &
         \epsilon_{\boldsymbol k}J_4(A) & 
      \\
      \vspace{.2cm}
       & -i\bar{\epsilon}_{\boldsymbol k}J_1(A) & \epsilon_{\boldsymbol k}J_0(A) & 
          i\bar{\epsilon}_{\boldsymbol k}J_1(A) & \epsilon_{\boldsymbol k}J_2(A) & 
          i\bar{\epsilon}_{\boldsymbol k}J_3(A) & 
      \\
      \vspace{.2cm}
      \cdots & \epsilon_{\boldsymbol k}J_2(A) & -i\bar{\epsilon}_{\boldsymbol k}J_1(A) & 
               \epsilon_{\boldsymbol k}J_0(A) &  i\bar{\epsilon}_{\boldsymbol k}J_1(A) & 
               \epsilon_{\boldsymbol k}J_2(A) &\cdots 
      \\
      \vspace{.2cm}
       & -i\bar{\epsilon}_{\boldsymbol k}J_3(A) & \epsilon_{\boldsymbol k}J_2(A) & 
         -i\bar{\epsilon}_{\boldsymbol k}J_1(A) & \epsilon_{\boldsymbol k}J_0(A) & 
          i\bar{\epsilon}_{\boldsymbol k}J_1(A) & 
      \\
       & \epsilon_{\boldsymbol k}J_4(A) & -i\bar{\epsilon}_{\boldsymbol k}J_3(A) & 
         \epsilon_{\boldsymbol k}J_2(A) & -i\bar{\epsilon}_{\boldsymbol k}J_1(A) & 
         \epsilon_{\boldsymbol k}J_0(A) &  
      \\
       & & &\vdots & & & \ddots
		\end{pmatrix},
	\label{ac_matrix}
\end{equation}
\end{widetext}
where we omit the label ``sc'' attached to $\epsilon_{\boldsymbol k}$ 
and $\bar{\epsilon}_{\boldsymbol k}$ for simplicity. 
Taking the dc limit $\Omega \to 0$ requires a special care 
because of definition (\ref{ac_a}) which is singular 
at $\Omega=0$. Therefore 
a quantity calculated for a system in the presence of the ac field with 
a finite frequency $\Omega\neq 0$ does not necessarily reproduce the 
result calculated for a system with the dc field discussed in Sec.~\ref{hc_dc}.

Let us examine the physical meaning of the Floquet matrix [Eq.~(\ref{ac_matrix})].
The $(m, n)$ component of Hamiltonian (\ref{ac_matrix}) is 
proportional to $J_{m-n}(A)$. Since $J_{m-n}(A)\propto A^{|m-n|}$ if $A$
is sufficiently small, the transition probability $m\to n$ is 
proportional to $E^{2|m-n|}$. For $m<n$, the process corresponds to 
stimulated absorption, while it corresponds to stimulated emission for $m>n$.
The process of spontaneous emission is not included since we assume that 
the electric field is classical so that there is no quantum fluctuation of 
photon numbers. This assumption is appropriate as long as the intensity 
of the electric field considered here be so strong as a pulsed laser. 

\subsection{Application to other lattices}
\label{other_lattice}

So far we have assumed that the vector potential ${\boldsymbol A}$ points to 
the specific direction $(1, 1, \dots,1)$ in the hypercubic lattice. 
We can more generally calculate the inverse of the noninteracting retarded Green's 
function ${G^{R0}}^{-1}$ by making use of formula (\ref{inv_g}) 
for arbitrary lattice structures and the vector potentials. 
The disadvantage of such a general case is that, 
since the ${\boldsymbol k}$ dependence of ${G^{R0}}^{-1}$ is not so simple as 
to be only through $\epsilon_{\boldsymbol k}^{\rm sc}$ and 
$\bar{\epsilon}_{\boldsymbol k}^{\rm sc}$, 
the integral over ${\boldsymbol k}$ becomes computationally heavier. 

Here we note that there is a group of lattice models that give 
a simple ${\boldsymbol k}$-dependence of ${G^{R0}}^{-1}$ in infinite dimensions. 
Among them are the face-centered cubic (fcc) (Sec.~\ref{fcc})
and body-centered cubic (bcc) (Sec.~\ref{bcc}) lattices with 
${\boldsymbol A}$ parallel to $(1, 1, \dots, 1)$
in infinite dimensions. 
It is instructive to give examples other than the hypercubic lattice.\\
%

\subsubsection{fcc lattice in infinite dimensions}
\label{fcc}

The energy dispersion of the fcc lattice generalized to arbitrary $d$ 
dimensions ($d\geq 2$) is given by 

\begin{equation}
  \epsilon_{\boldsymbol k}^{\rm fcc}
    =
      \frac{4}{2\sqrt{d(d-1)}}\sum_{\alpha=2}^d \sum_{\beta=1}^{\alpha-1}
      \cos k_\alpha \cos k_\beta.
  \label{fcc_infinite}
\end{equation}
In the infinite-dimensional limit ($d\to\infty$) the dispersion of the 
fcc lattice $\epsilon_{\boldsymbol k}^{\rm fcc}$ is related to that for 
the sc lattice [Eq.~(\ref{e_sc})] through,\cite{muller-hartmann1989a} 
\begin{equation}
  \epsilon_{\boldsymbol k}^{\rm fcc}
    =
      (\epsilon_{\boldsymbol k}^{\rm sc})^2-\frac{1}{2}.
  \label{fcc_sc}
\end{equation}
Using Eqs.~(\ref{inv_g}) and (\ref{fcc_sc}), we derive the inverse of 
Green's function for the infinite-dimensional fcc lattice in the 
dc field, 
\begin{align}
  ({G_{\boldsymbol k}^{R0}}^{-1})_{mn}(\omega)
    &=
      (\omega+n\Omega+\mu+i\eta)\delta_{mn}
  \nonumber
  \\
    &\quad 
		  -\frac{1}{4}e^{i(m-n)\theta_{\boldsymbol k}}
				[
           \zeta_{\boldsymbol k}^2 (\delta_{m-n,2} + \delta_{m-n,-2})
	\nonumber
	\\
	  &\quad
          +2(\zeta_{\boldsymbol k}^2-1)\delta_{mn}
        ].
	\label{fcc_dc}
\end{align}
We see that the Hamiltonian part of the inverse of Green's function [the second 
term on the rhs of Eq.~(\ref{fcc_dc})] is written in a pentadiagonal matrix form. 
In the same way we can obtain Green's function on the ac field via Eq. 
(\ref{inv_g}). The result reads
\begin{align}
  &({G_{\boldsymbol k}^{R0}}^{-1})_{mn}(\omega)
    =
      (\omega+n\Omega+\mu+i\eta)\delta_{mn}
  \nonumber
  \\
    &
		  -\frac{1}{2}
        \begin{cases}
           \zeta_{\boldsymbol k}^2 \cos(2\theta_{\boldsymbol k}) J_{m-n}(2A)
          +(\zeta_{\boldsymbol k}^2-1)\delta_{mn}
          & m-n{\rm : even}
          \\
          i\, \zeta_{\boldsymbol k}^2 \sin(2\theta_{\boldsymbol k}) J_{m-n}(2A)
          & m-n{\rm : odd}
        \end{cases}.
	\label{fcc_ac}
\end{align}
We notice that the factors $J_{m-n}(2A)$ and $J_{m-n}(0)=\delta_{mn}$ appear 
in Eq.~(\ref{fcc_ac}) [while $J_{m-n}(A)$ appears on the sc 
lattice; see Eq.~(\ref{sc_ac})]. Equations (\ref{fcc_dc}) and 
(\ref{fcc_ac}) indicate that Green's functions depend on ${\boldsymbol k}$ 
only via the two functions $\epsilon_{\boldsymbol k}^{\rm sc}$ [Eq.~(\ref{e_sc})] and 
$\bar{\epsilon}_{\boldsymbol k}^{\rm sc}$ [Eq.~(\ref{ebar_sc})], so that 
we can integrate over ${\boldsymbol k}$ using the JDOS [Eq.~(\ref{jdos_def})].

\subsubsection{bcc lattice in infinite dimensions}
\label{bcc}

The bcc lattice in $d$ dimensions ($d\ge 3$) is defined by 
the dispersion relation,
\begin{equation}
  \epsilon_{\boldsymbol k}^{\rm bcc}
    =
      -\frac{8}{2\sqrt{d(d-1)(d-2)}}\sum_{\alpha=3}^d 
			\sum_{\beta=2}^{\alpha-1} \sum_{\gamma=1}^{\beta-1}
      \cos k_\alpha \cos k_\beta \cos k_\gamma.
\end{equation}
We can take the limit $d\to\infty$ in the same way as 
in the case of the fcc lattice, and the dispersion converges to 
\begin{equation}
  \epsilon_{\boldsymbol k}^{\rm bcc}
    =
      \frac{2}{3}(\epsilon_{\boldsymbol k}^{\rm sc})^3-\epsilon_{\boldsymbol k}^{\rm sc},
  \label{bcc_sc}
\end{equation}
from which we can derive Green's function in the dc field, 
\begin{widetext}
\begin{align}
  ({G_{\boldsymbol k}^{R0}}^{-1})_{mn}(\omega)
    &=
      (\omega+n\Omega+\mu+i\eta)\delta_{mn}
		  -\frac{1}{12}e^{i(m-n)\theta_{\boldsymbol k}}\zeta_{\boldsymbol k}
			[
        \zeta_{\boldsymbol k}^2 (\delta_{m-n,3}+\delta_{m-n,-3})
        +3(\zeta_{\boldsymbol k}^2-2) (\delta_{m-n,1}+\delta_{m-n,-1})
      ].
	\label{bcc_dc}
\end{align}
In this case the Hamiltonian part of the inverse of Green's function 
[the second term on the rhs of Eq.~(\ref{bcc_dc})] becomes a 
heptadiagonal matrix. Similarly, Green's function in the ac field is 
written as 
\begin{align}
  ({G_{\boldsymbol k}^{R0}}^{-1})_{mn}(\omega)
    &=
      (\omega+n\Omega+\mu+i\eta)\delta_{mn}
		  -\frac{1}{6}
        \begin{cases}
            \displaystyle
						\zeta_{\boldsymbol k}
            [
					    \zeta_{\boldsymbol k}^2\cos(3\theta_{\boldsymbol k})J_{m-n}(3A)
              +3(\zeta_{\boldsymbol k}^2-2)\cos\theta_{\boldsymbol k}J_{m-n}(A)
						]
            & m-n{\rm : even}
          \\
            \displaystyle
            i\, \zeta_{\boldsymbol k}
						[
							\zeta_{\boldsymbol k}^2 \sin(3\theta_{\boldsymbol k})J_{m-n}(3A)
              +3(\zeta_{\boldsymbol k}^2-2) \sin\theta_{\boldsymbol k}J_{m-n}(A)
						]
            & m-n{\rm : odd}
        \end{cases},
	\label{bcc_ac}
\end{align}
\end{widetext}
where the factor $J_{m-n}(3A)$ newly appears besides $J_{m-n}(A)$. 
Again, Green's functions depend on ${\boldsymbol k}$ only through 
$\epsilon_{\boldsymbol k}^{\rm sc}$ and $\bar{\epsilon}_{\boldsymbol k}^{\rm sc}$, 
which makes an integral over ${\boldsymbol k}$ computationally quite efficient with the JDOS 
[Eq.~(\ref{jdos_def})]. 

\section{Diagonalization of Floquet matrices}
\label{diagonalization}

Here we examine how to diagonalize a Floquet matrix 
$(\omega+n\Omega)\delta_{mn}-H_{mn}$ appearing in the 
original Schr\"{o}dinger Eq.~(\ref{floquet}) for the 
noninteracting electrons. 
We have shown in Sec.~\ref{general_lattice_field} that the Floquet matrix 
form of Green's function is diagonalized into $Q_{\boldsymbol k}(\omega)$
by the unitary transformation 
$\Lambda_{\boldsymbol k}$ [see Eq.~(\ref{lambda_q_lambda})]. 
In Sec.~\ref{inverse_green}, we have mentioned that the inverse of 
Green's function is equivalent to the Floquet matrix 
$(\omega+n\Omega+\mu+i\eta)\delta_{mn}-H_{mn}$. Combining these two facts, 
we identify the eigenvalues and the eigenvectors of the Floquet matrix. 
We summarize the statement below.
{\it The eigenvalues of a Floquet matrix 
$H_{mn}-n\Omega\delta_{mn}$ 
for a single-band noninteracting system 
subject to a homogeneous electric field periodic in time 
are }
\begin{align}
  -(Q_{\boldsymbol k}^{-1})_{nn}(0)
    &=
	    (\epsilon_{\boldsymbol k})_0 - n\Omega 
	\nonumber
  \\
    &=
	    H_{nn} - n\Omega \quad
		  (n=0,\pm 1,\pm 2, \dots),
	\label{Shirley}
\end{align}
{\it and for each $n$, the corresponding eigenvector is given by}
\begin{equation}
  u_{\boldsymbol k}^{m-n}
	  =
		  (\Lambda_{\boldsymbol k})_{mn}
			\quad (m=0,\pm 1,\pm 2, \dots).
  \label{theorem}
\end{equation}
This completely solves Floquet matrix problems for 
a single-band system of noninteracting electrons. 
We note that the fact that Eq.~(\ref{Shirley}) gives the eigenvalues is 
essentially a consequence of Shirley's relation [Eq.~(5) of Ref.~\onlinecite{shirley1965}]. 
Since $(\epsilon_{\boldsymbol k})_0\tau$ is a dynamical phase, 
the above statement indicates the absence of an additional geometrical 
phase (see Appendix \ref{inverse_g}). 

From Eqs.~(\ref{lambda_q_lambda}) and (\ref{theorem}), the 
noninteracting Green's function can be written with the Floquet states as 
\begin{align}
  (G_{\boldsymbol k}^{R0})_{mn}(\omega)
	  =
		  \sum_\ell \frac{u_{\boldsymbol k}^{m-\ell}(u_{\boldsymbol k}^{n-\ell})^\ast}
			{\omega+\ell\Omega+\mu-(\epsilon_{\boldsymbol k})_0+i\eta}.
\end{align}
One should note that the theorem cannot be applied to a 
multiband system, where contributions of a geometrical phase may survive 
and interband transition could be caused by the electric field. 

\begin{figure}[t]
  \begin{center}
    \includegraphics[width=7cm]{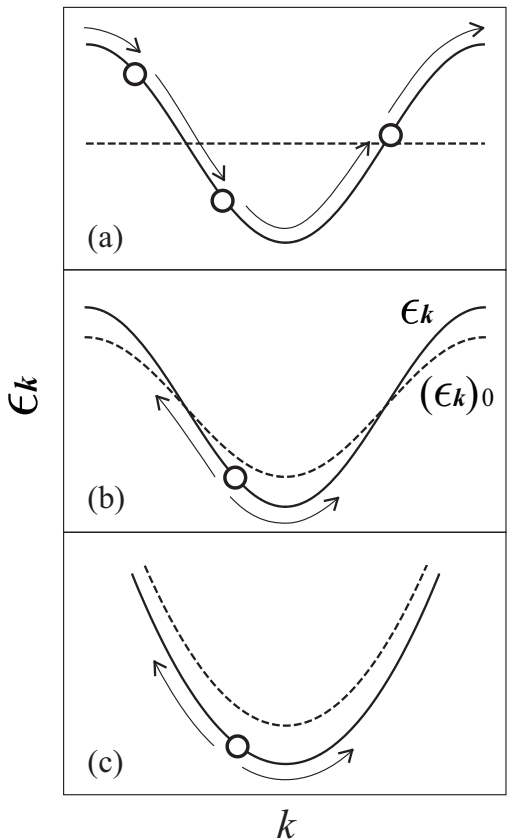}
		\caption{The band renormalization $\epsilon_{\boldsymbol k} \to 
		(\epsilon_{\boldsymbol k})_0$ on the (hyper)cubic lattice in (a) the dc field and (b) the ac field
		and (c) on a parabolic band in the ac field.
		The solid line represents $\epsilon_{\boldsymbol k}$, while the dashed 
    line respresents $(\epsilon_{\boldsymbol k})_0$. The circles are occupied states 
		moving along the arrows. }
		\label{band_renormalization}
  \end{center}
\end{figure}

The theorem provides 
a unified description of periodically driven systems: the original band 
structure $\epsilon_{\boldsymbol k}$ is renormalized into the time-averaged 
one $(\epsilon_{\boldsymbol k})_0$ due to the field, 
and the renormalized band splits into its {\it replicas} with the spacing $\Omega$. 
On the (hyper)cubic lattice, 
the dc field changes the band into the Wannier-Stark ladder with 
the infinitesimal bandwidth [Fig.~\ref{band_renormalization}(a)], 
while in the ac field, the bandwidth 
scales with the factor $J_0(A)$ [Fig.~\ref{band_renormalization}(b)]. 

To see how the theorem actually works, 
let us apply it to the $d$-dimensional fcc lattice model in the dc field as an example. 
For convenience, we restrict our discussion in the limit $d \to \infty$. 
The Floquet matrix form of the 
Hamiltonian is given by Eq.~(\ref{fcc_dc}). According to the 
theorem, its eigenvalues are equal to the diagonal components: 
$\omega+n\Omega-\frac{1}{2}(\zeta_{\boldsymbol k}^2-1)$. This means that the 
dc field modifies the energy dispersion 
from $\epsilon_{\boldsymbol k}^{\rm fcc}$ [Eq.~(\ref{fcc_sc})]
to $\tilde{\epsilon}_{\boldsymbol k}^{\rm fcc}\equiv\frac{1}{2}(\zeta_{\boldsymbol k}^2-1)$. 
The latter dispersion is equivalent to the one for a 
$(d-1)$-dimensional hyperplane perpendicular to $(1, 1, \dots, 1)$, 
the direction of the electric field. 
We can interpret this fact as follows: the dc field 
makes the electrons localize along the direction of the field 
due to the Bloch oscillation, but the electrons are free to move along 
the directions perpendicular to the field because along those directions 
the force of the dc field does not act on the electrons. 
As a result, the motion of the electrons is confined on the hyperplane, and the 
energy dispersion becomes $\tilde{\epsilon}_{\boldsymbol k}^{\rm fcc}$.

Another example is provided by an ac field. 
If one sees the Floquet matrix form of the Hamiltonian for the system in an ac field 
on the fcc [Eq.~(\ref{fcc_ac})] or bcc [Eq.~(\ref{bcc_ac})] lattices, 
one finds that various kinds of band renormalization besides the 
$J_0(A)$ scaling on the simple cubic lattice are implied by the theorem. 

Finally, let us consider electrons in a parabolic band 
(e.g., a conduction-band bottom in a semiconductor) 
with $\epsilon_{\boldsymbol k}={\boldsymbol k}^2/2m^\ast$.
If we apply an ac field on it, the dispersion is renormalized into 
$(\epsilon_{\boldsymbol k})_0=\epsilon_{\boldsymbol k}+e^2E^2/4m^\ast\Omega^2$
[Fig.~\ref{band_renormalization}(c)].
This results from the dynamical Franz-Keldysh effect,\cite{jauho_johnsen1996, nordstrom_jajbknas1998} 
which is, in the present formalism, naturally understood through the diagonalization picture of the Floquet matrix. 

\section{Dynamical mean-field theory with the Floquet-Green function method}
\label{dmft}

Now we are in position to combine FGFM with DMFT for treating 
interacting systems in external fields. 
At the basis of DMFT lies the fact that a lattice problem of correlated 
many-body systems can be approximately mapped to a problem of an impurity 
embedded into the environment of an effective medium when one ignores 
spatial fluctuations but takes fully into account on-site 
dynamical correlation.\cite{georges_kotliar1992, georges_kkr1996} The mapping is given as follows: let
$Z=\int [dc_i][dc_i^\dagger]\; e^{iS[c_i, c_i^\dagger]}$ be a partition 
function in terms of the original action, 
$S=\int dt \int dt' \sum_{ij} c_i^\dagger(t) {G_{ij}^0}^{-1}(t, t')c_j(t')
+\sum_iS_{\rm int}[c_i, c_i^\dagger]$, where the interaction term is 
assumed to be a sum of the local terms. Integrating out each site's 
degrees of freedom except for a representative site $i=o$, 
we have the local partition function, 
$Z_{\rm loc}[\mathscr{G}_0]=\int [dc_o][dc_o^\dagger]\; e^{iS_{\rm loc}[c_o, c_o^\dagger]}$.
Here the local action reads
$S_{\rm loc}=\int dt \int dt' c_o^\dagger(t)\mathscr{G}_0^{-1}(t,t')c_o(t')
+S_{\rm int}[c_o, c_o^\dagger]$ [$\mathscr{G}_0(t,t')$: the Weiss 
function]. If one ignores the nonlocal fluctuations, 
$Z$ and $Z_{\rm loc}$ give 
a common site-diagonal self-energy, $\Sigma_{ij}(t,t')=\delta_{ij}\Sigma(t,t')$. 
This fact enables us to build a set of self-consistent closed equations, 
which can be solved with an iterative numerical calculation. 
Although neglecting the spatial fluctuation is generally an 
approximation, the nonlocal corrections become rigorously irrelevant 
in the limit of infinite dimensions, where the 
hopping parameter is scaled as $t=t^\ast/2\sqrt{d}$ 
($t^\ast$: fixed).\cite{metzner_vollhardt1989} 

DMFT is also applicable to nonequilibrium systems as recently studied
\cite{freericks_tz2006, eckstein_kollar2008, schmidt_monien2002} based on nonequilibrium 
Green's-function formalism. 
Similar to the equilibrium case, one has self-consistent equations 
for Green's function and the self-energy. 
Now, the present proposal is that 
if the driven system is periodic in time, one can rewrite the equations 
in the Floquet matrix form, 
\begin{align}
  (G_{\rm loc})_{mn}(\omega)
	  &=
	    \frac{\delta Z_{\rm loc}[\mathscr{G}_0]}{\delta (\mathscr{G}_0^{-1})_{nm}(\omega)},
	\label{self-consistent1}
  \\
  (G_{\rm loc}^{-1})_{mn}(\omega)
    &=
	    (\mathscr{G}_0^{-1})_{mn}(\omega) - \Sigma_{mn}(\omega),
	\label{self-consistent2}
  \\
  (G_{\boldsymbol k}^{-1})_{mn}(\omega)
    &=
	    ({G_{\boldsymbol k}^0}^{-1})_{mn}(\omega) - \Sigma_{mn}(\omega),
	\label{self-consistent3}
  \\
  (G_{\rm loc})_{mn}(\omega)
    &=
	    \sum_{\boldsymbol k} (G_{\boldsymbol k})_{mn}(\omega).
	\label{self-consistent4}
\end{align}
To solve Eqs.~(\ref{self-consistent1})-(\ref{self-consistent4}) 
self-consistently, we first input the inverse of the noninteracting 
Green's function given by Eq.~(\ref{inv_g}) into Eq.~(\ref{self-consistent3}).
After the initial self-energy is properly chosen, 
the calculation is iterated until Green's function converges. 
For the lattices discussed in Secs.~\ref{hypercubic} and \ref{other_lattice}, 
the integral over ${\boldsymbol k}$ in Eq.~(\ref{self-consistent4})
is performed via the JDOS [Eq.~(\ref{jdos_def})]. 

As remarked in Sec.~\ref{flq_rep_g}, the size of the Floquet matrix that 
needs to be taken in a calculation is usually small ($\sim$ 5-30, 
depending on $\Omega$), 
which, with the analytic expression of the inverse of Green's function 
(\ref{inv_g}), makes our computational costs dramatically small. 

\section{Gauge-invariant Green's function}
\label{gauge}

Before applying our method to a model, we examine the gauge invariance 
of Green's function. Let us write the coordinates 
$x^\nu = (t, {\boldsymbol r})$ and the vector potential 
$A^\nu = (\phi, {\boldsymbol A})$ in the four-vector form. 
The gauge transformation, 
$A_\nu(x) \to A_\nu(x) + \partial_\nu \chi(x)$, puts a phase factor 
to the creation and the annihilation operators as  
$c^\dagger(x) \to e^{-ie\chi(x)}c^\dagger(x)$ and $c(x) \to 
e^{ie\chi(x)}c(x)$. Accordingly Green's function changes 
as $G(x, x') \to e^{ie[\chi(x)-\chi(x')]}G(x, x')$, i.e., the 
usual Green's function is not gauge invariant. It is 
known\cite{boulware1966, turkowski_freericks2006} that one can 
make Green's function gauge invariant with an additional phase factor as 
\begin{equation}
  \tilde{G}(x, x')
	  =
		  \exp\left(
			  -i\int_{x'}^{x} dy^{\nu} \;
			eA_{\nu}(y)
			\right)
			G(x, x'). 
\end{equation}
$\tilde{G}$ depends on the path of the line integral in the 
exponent. Here we adopt the conventional straight line 
connecting $x$ with $x'$ as the path of the integral. 
Suppose that Green's function depends on ${\boldsymbol k}$ only through 
$\epsilon_{\boldsymbol k}^{\rm sc}$ and $\bar{\epsilon}_{\boldsymbol k}^{\rm sc}$. 
Then in the temporal gauge ($\phi=0$) 
the Wigner representation of the modified Green's function $\tilde{G}$ becomes 
\begin{align}
  \tilde{G}_m(\zeta, \theta, \omega)
	  &=
		  \sum_n \int \frac{d\omega '}{2\pi}\; \frac{1}{\tau}
			\int_{-\tau/2}^{\tau/2} dt_{\rm av} \int dt_{\rm rel}\;
	\nonumber
	\\
			&\quad \times e^{i(m-n)\Omega t_{\rm av}+i(\omega-\omega ')t_{\rm rel}}
	\nonumber
	\\
	  &\quad \times
			G_n\left(
			  \zeta, \theta+\int_{-1/2}^{1/2}d\lambda\;
				eA(t_{\rm av}+\lambda t_{\rm rel}), \omega '
			\right),
	\label{tilde}
\end{align}
where $\zeta$ and $\theta$ are defined in Eqs.~(\ref{zeta_theta1}) and (\ref{zeta_theta2}). 
Note that $\tilde{G}$ is calculated by shifting the variable 
$\theta$ in the original Green's function. 
This suggests that 
Green's function integrated in terms of $\theta$ is definitely gauge invariant, so that 
the local Green's function $\sum_{\boldsymbol k}G_{\boldsymbol k}(\omega)$ is also 
gauge invariant.\cite{turkowski_freericks2006} 
In the following, we derive the gauge-invariant Green's function 
$\tilde{G}$ for the dc field in Sec.~\ref{gauge_dc} and 
the ac field in Sec.~\ref{gauge_ac}. 

\subsection{dc field}
\label{gauge_dc}

To obtain $\tilde{G}$ for the dc field, we first note that the Hamiltonian 
in the dc field [Eq.~(\ref{dc_vector_p})] has the time translation symmetry. If one makes a 
time translation $t \to t+\delta t$, the vector potential changes as 
$A(t) \to A(t) + \Omega\delta t$. Since the change can be absorbed by the 
gauge transformation with $\chi=-\Omega \delta t\sum_{i=1}^d x^i$, the Hamiltonian 
is invariant against time translation. 
Then we {\it assume} that in the long-time limit after the dc field is switched on 
the retarded Green's function becomes independent of the initial correlations. 
This assumption seems to be valid\cite{tran2008_} as numerically checked in Sec.~\ref{fk_dc}. 
As a result, a gauge-invariant quantity that is calculated from the 
retarded Green's function should be independent of the average time 
$t_{\rm av}$. 
For instance, the local Green's function in the Wigner representation 
$(G_{\rm loc}^R)_n(\omega)$, which is gauge invariant as shown above,
vanishes for $n\neq 0$, so that $G_{\rm loc}^R(t, t')$ does not depend on 
$t_{\rm av}$. In the same way the self-energy $\Sigma_n^R(\omega)$ also vanishes for 
$n\neq 0$. 

Since the Floquet-represented self-energy $\Sigma_{mn}^R(\omega)$ is 
diagonal due to the symmetry, we can identify the 
$\theta$ dependence of the Floquet representation of the retarded Green's function as
\begin{equation}
  G_{mn}^R(\zeta, \theta, \omega)
	  =
		  e^{i(m-n)\theta} G_{mn}^R(\zeta, \theta=0, \omega).
	\label{theta_dep}
\end{equation}
Using Eq.~(\ref{theta_dep}), one can evaluate the gauge-invariant Green's 
function (\ref{tilde}) as 
\begin{equation}
  \tilde{G}_m^R(\zeta, \theta, \omega)
	  =
		  \delta_{m, 0} \sum_n G_n^R(\zeta, \theta, \omega), 
	\label{dc_gauge}
\end{equation}
where we can see that every mode of Green's function {\it equally} contributes to $\tilde{G}^R$. 

\subsection{ac field}
\label{gauge_ac}

For the ac field, which is one of the key questions in the present paper, 
Green's function has a more complicated $\theta$ dependence. 
To evaluate the gauge-invariant Green's function (\ref{tilde}), here we 
expand the original Green's function with respect to $eA(t)$ 
in a Taylor series: $G_n^R(\zeta, \theta+\int d\lambda\; eA, \omega)=\sum_\ell
\frac{1}{\ell !}(\int d\lambda\; eA)^\ell\;
\partial_\theta^\ell G_n^R(\zeta,\theta,\omega)$. 
Then the Wigner representation of the gauge-invariant Green's function is 
expressed as 
\begin{align}
  \tilde{G}_m^R(\zeta, \theta, \omega)
	  &=
		  \sum_{\ell n} \frac{2 A^\ell}{\ell ! \Omega} \int d\omega ' \;
			\partial_\theta^\ell G_n^R(\zeta, \theta, \omega ') \; 
		\nonumber
	\\
      &\quad \times X_{m-n}^{(\ell)} \;
			Y^{(\ell)}\!\left(\frac{2}{\Omega}(\omega-\omega ')\right),
	\label{ac_gauge}
\end{align}
where
\begin{align}
  X_{m-n}^{(\ell)}
	  &\equiv 
		  \int_{-\pi}^{\pi} \frac{dx}{2\pi} \;
			e^{i(m-n)x} \sin^\ell x
	\nonumber
	\\
		&=
		  \frac{1}{(2i)^\ell} \sum_{r=0}^{\ell}
			\binom{\ell}{r}
			(-1)^r \; \delta_{m-n, 2r-\ell},
\end{align}
and
\begin{align}
	&Y^{(\ell)}(k)
	  \equiv 
		  \int_{-\infty}^{\infty} \frac{dx}{2\pi} \; e^{ikx} \left(\frac{\sin x}{x}\right)^\ell
	\nonumber
	\\
		&=
		  \frac{1}{2^\ell (\ell-1)!} \sum_{r=0}^\ell
			\binom{\ell}{r}
			(-1)^r (k+\ell-2r)^{\ell-1} \theta(k+\ell-2r)
\end{align}
for $\ell\ge 1$.
When $\ell=0$, we have $Y^{(\ell)}(k) = \delta(k)$. 
The Floquet representation of Eq.~(\ref{ac_gauge}) reads 
\begin{align}
  &\tilde{G}_{mn}^R(\zeta, \theta, \omega)
	  =
		  G_{mn}^R(\zeta, \theta, \omega) +
		  \sum_{\ell=1}^\infty \frac{2}{\ell ! \Omega} \left(\frac{A}{2i}\right)^\ell
	\nonumber
	\\
			&\times \sum_{r=0}^{\ell} (-1)^r
			\binom{\ell}{r}
			\left(
			  \int_{\omega}^{\Omega/2} d\omega ' \sum_{k=0}^{\ell-1}
				+\int_{-\Omega/2}^{\omega} d\omega ' \sum_{k=1}^{\ell}
			\right)
	\nonumber
	\\
	  &\times
			\partial_\theta^\ell G_{m-r+k, n+r+k-\ell}^R(\zeta, \theta, \omega ')
			Y^{(\ell)}\!\left(\frac{2}{\Omega}(\omega-\omega ')-2k+\ell\right).
	\label{ac_gauge_flq}
\end{align}
The derivative with respect to $\theta$ in Eqs.~(\ref{ac_gauge}) and 
(\ref{ac_gauge_flq}) is 
simplified when the system is on the hypercubic lattice. 
The Floquet representation of the noninteracting Hamiltonian $H$ is then given by 
Eq.~(\ref{ac_matrix}), and the $\ell$th derivative of the Floquet representation of the retarded Green's 
function can be calculated for every $\ell$ via the recurrence relations 
\begin{align}
  \partial_\theta H
    &=
	    \bar{H},
	\label{recurr1}
 	\\
  \partial_\theta \bar{H}
    &=
	    -H,
	\label{recurr2}
 	\\
  \partial_\theta G^R
    &=
	    -G^R (-\partial_\theta H) G^R
 		=
  	  G^R \bar{H} G^R.
	\label{recurr3}
\end{align}
Employing Eqs.~(\ref{recurr1})-(\ref{recurr3}) with relation 
(\ref{ac_gauge_flq}), one can numerically evaluate 
the gauge-invariant retarded Green's function $\tilde{G}^R$ for the ac 
field. 

\section{Application to the Falicov-Kimball model}
\label{fk}

To test the ability of the present method for treating many-body systems, 
we apply it to the spinless FK model, for which the Hamiltonian is 
\begin{equation}
  H
	  =
		  - \sum_{ij} t_{ij} c_i^\dagger c_j
			+ U \sum_i c_i^\dagger c_i f_i^\dagger f_i.
\end{equation}
Here $f_i$ ($f_i^\dagger$) annihilates (creates) a localized electron, and $U$ is 
a coupling constant. It is known that the FK model exhibits a metal-insulator transition in 
infinite dimensions from DMFT calculations, where the possibility of 
charge-density wave phases is ignored.\cite{freericks_zlatic2003} The critical value of $U$ 
for the transition is known 
to be $\sqrt{2}$ on the hypercubic lattice at half filling. 
The insulating phase is Mott-like, which means that the insulating state 
originates from the electron correlation. 

What characterizes the FK model is that it has an exact solution for 
the impurity problem [Eq.~(\ref{self-consistent1})] within DMFT, even out 
of equilibrium.\cite{freericks_tz2006} The solution for the retarded 
Green's function is 
\begin{equation}
  G_{\rm loc}^R(\omega)
	  =
	    w_0 \;\mathscr{G}_0^R(\omega) +
			w_1 \;[{\mathscr{G}_0^R}^{-1}(\omega) - U]^{-1},
\end{equation}
where $w_1$ is the filling of the $f$ electrons, and $w_0=1-w_1$. Note 
that the retarded component of Green's function decouples to the Keldysh component. 
Here we concentrate on the retarded Green's function, and 
calculate the local spectral function $A_n(\omega)=-\frac{1}{\pi}{\rm Im}
(G_{\rm loc}^R)_n(\omega)$ (which is gauge invariant as explained in 
Sec.~\ref{gauge}) and 
the gauge-invariant spectral function $\tilde{A}_n({\boldsymbol k}, \omega)=-\frac{1}{\pi}
{\rm Im}(\tilde{G}_{\boldsymbol k}^R)_n(\omega)$ under the assumption 
that the initial correlations are irrelevant to the retarded Green's function.


\subsection{Falicov-Kimball model in the dc field}
\label{fk_dc}

We first present the results for the dc field. 
We start with noting that the integral over $\theta$ can be performed 
analytically due to relation (\ref{theta_dep}), and that $G_{\rm loc}^R$, 
$\mathscr{G}_0^R$, and $\Sigma^R$ in the Floquet representation 
are all diagonal as mentioned in Sec.~\ref{gauge_dc}, 
which simplifies our calculation. 
All the matrices we have to invert numerically 
are tridiagonal owing to Eq.~(\ref{sc_dc1}). 

\begin{figure}[t]
  \begin{center}
    \includegraphics[width=8.5cm]{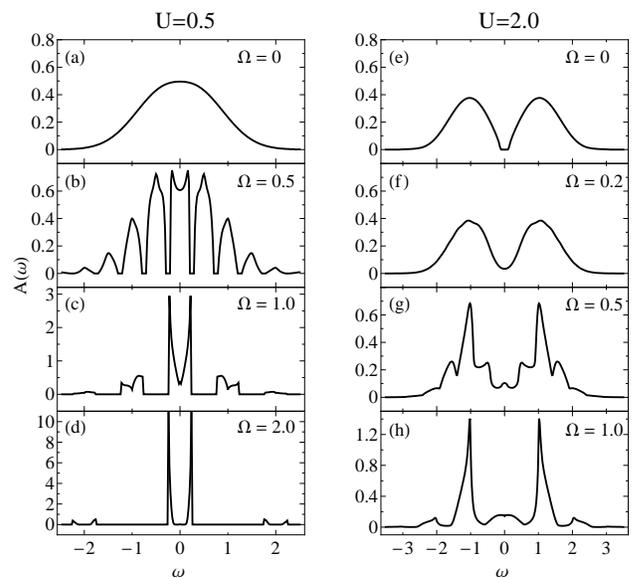}
    \caption{The local spectral function $A_0(\omega)$ for the FK 
    model coupled to the dc field on the hypercubic lattice at half 
    filling. }
    \label{data_dc}
  \end{center}
\end{figure}
In Fig.~\ref{data_dc} we illustrate the local spectral function 
$A_0(\omega)$ for various values of $U$ and $\Omega$ on the hypercubic lattice. 
The size of the Floquet matrices that we choose is typically 9-13.
Convergence is achieved after typically 10-30 iterations, 
where the calculation is quite stable over the parameter space considered here. 
We can see that the present result in Figs.~\ref{data_dc}(b)-\ref{data_dc}(d) 
obtained in the Floquet method agrees well 
with the previous results,\cite{freericks2008, turkowski_freericks2006} 
where the nonequilibrium DMFT is employed. 
This suggests that our assumption of the irrelevance of the initial 
correlations to the retarded Green's function is valid.
In our results, 
the spectral function is positive definite, 
and satisfies the sum rule for the zeroth spectral moment\cite{turkowski_freericks2006a} 
as in equilibrium. Therefore we can safely
interpret the quantity $A(\omega)$ as the spectrum of the system even 
out of equilibrium. 

More interesting case is 
Figs.~\ref{data_dc}(e)-\ref{data_dc}(h), where we can observe how a Mott-like insulator 
($U=2$) is driven into a metallic state by the dc field. Namely, 
while there is a clear band gap between the 
upper and lower bands in equilibrium [Fig.~\ref{data_dc}(e), $\Omega = 0$], 
the gap disappears as the dc field is increased, where the spectral weight 
around $\omega = 0$ develops. Hence our calculation captures the Mott-like 
insulator-to-metal transition induced by a static electric field. 
In the strong dc field region [Figs.~\ref{data_dc}(b)-\ref{data_dc}(d), 
\ref{data_dc}(g), and \ref{data_dc}(h)], 
we find complicated structures with the spacing $\Omega$. We can  
attribute these to the Wannier-Stark ladder (mentioned in Sec.~\ref{hc_dc}),
which grows with the field intensity $E \propto \Omega$ [see Eq.~(\ref{dc_omega})]. 
The Wannier-Stark structure 
interferes with the original spectrum that comprises two bands with the spacing $U$, 
producing a characteristic interference pattern. 

\subsection{Falicov-Kimball model in the ac field}
\label{fk_ac}

\begin{figure}[t]
  \begin{center}
    \includegraphics[width=8.5cm]{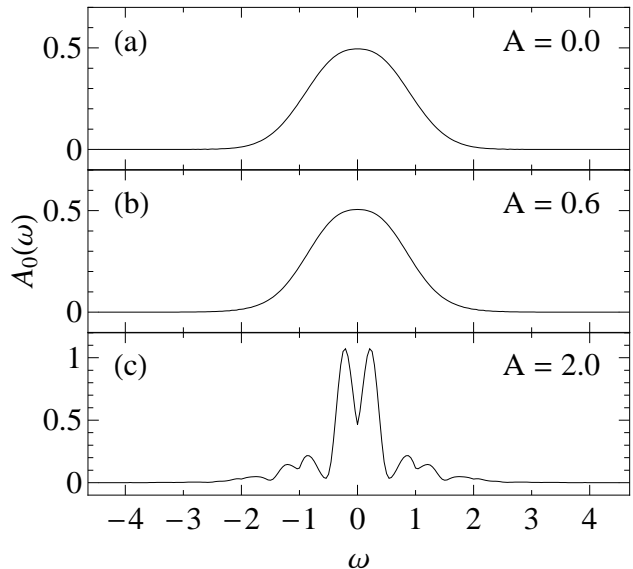}
		\caption{The zeroth mode of the local spectral function $A_0(\omega)$
		of the FK model coupled to the ac field ($\Omega=1$) on the hypercubic
		lattice at half filling for $U=0.5$. }
		\label{results_ac21}
  \end{center}
\end{figure}
\begin{figure}[t]
  \begin{center}
    \includegraphics[width=8.5cm]{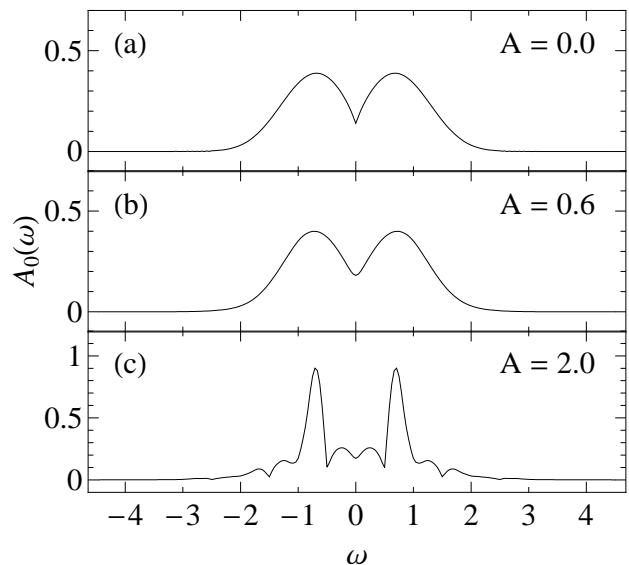}
		\caption{The zeroth mode of the local spectral function $A_0(\omega)$
		of the FK model coupled to the ac field ($\Omega=1$) on the hypercubic
		lattice at half filling for $U=1.3$. }
		\label{results_ac22}
  \end{center}
\end{figure}
\begin{figure}[t]
  \begin{center}
    \includegraphics[width=8.5cm]{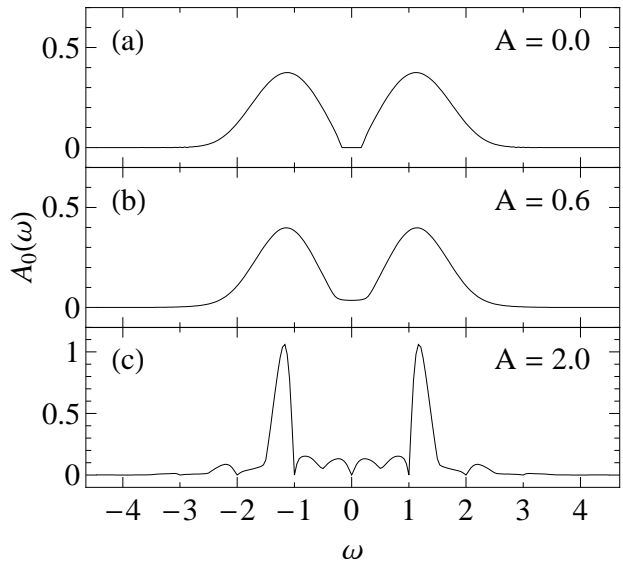}
		\caption{The zeroth mode of the local spectral function $A_0(\omega)$
		of the FK model coupled to the ac field ($\Omega=1$) on the hypercubic
		lattice at half filling for $U=2.2$. }
		\label{results_ac31}
  \end{center}
\end{figure}
\begin{figure}[t]
  \begin{center}
    \includegraphics[width=8.5cm]{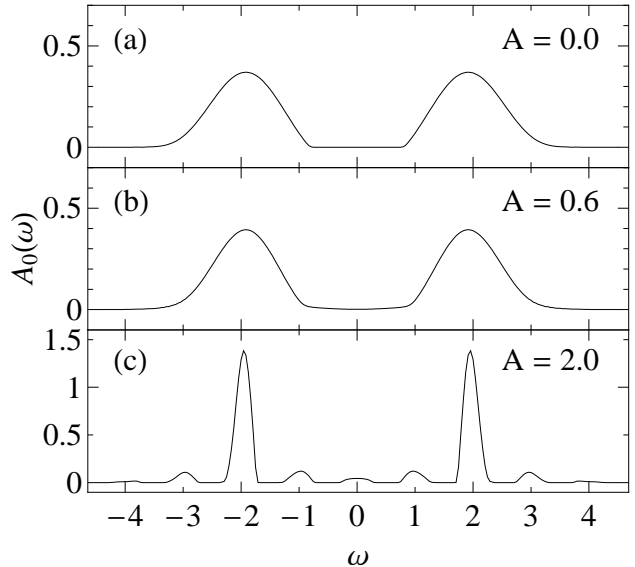}
		\caption{The zeroth mode of the local spectral function $A_0(\omega)$
		of the FK model coupled to the ac field ($\Omega=1$) on the hypercubic
		lattice at half filling for $U=3.8$. }
		\label{results_ac32}
  \end{center}
\end{figure}

We move on to the ac field. While the spectral function is 
time independent for the dc field, 
i.e., the non-zeroth modes of $A_n(\omega)$ vanish, 
this is no longer the case for the ac field. Since our interest resides in the 
time-averaged spectral function 
$\int dt_{\rm av}\; A(\omega, t_{\rm av}) = A_0(\omega)$, 
we concentrate on the zeroth mode of the spectral function. 
Unlike the dc case the integral over $\theta$ 
is nontrivial, which has to be calculated numerically. 
In Figs.~\ref{results_ac21}-\ref{results_ac32}, we depict the local 
spectral function $A_0(\omega)$ on the hypercubic lattice at half 
filling with the frequency of the ac field $\Omega=1$. 
The efficiency of the convergence and the 
stability of the calculation are similar to the dc case. 

In the metallic region (Fig.~\ref{results_ac21} for $U=0.5$ and 
Fig.~\ref{results_ac22} for $U=1.3$), one 
can see how the metallic spectrum of the system is deformed by the ac field. 
Namely, the width of the band shrinks with the intensity of the field. 
It can even goes to zero when $A$ coincides with a zero of $J_0(A)$, 
which makes interacting electrons localize. 
This is quite similar to the noninteracting case as examined in Sec.~\ref{hc_ac}. 
Hence we have {\it the dynamical localization in interacting 
electron systems}. Note that the scaling of the band width with $J_0(A)$
is a nonlinear effect of the ac field, as evident from $J_0(A)=1-(A/2)^2+\cdots$.
The difference between noninteracting and 
interacting cases is that each peak in the dynamical Wannier-Stark ladder 
at $\omega = n\Omega\; (n=0, \pm 1, \pm 2, \dots)$
splits into two with the spacing $U$ due to the correlation effect. 
This can clearly be seen in Fig.~\ref{results_ac21}(c). 

In the insulating region (Fig.~\ref{results_ac31} for $U=2.2$ and 
Fig.~\ref{results_ac32} for $U=3.8$), 
on the other hand, we do observe the ac-field driven transition 
from the Mott-like insulating state to a metallic state. 
In equilibrium [Figs.~\ref{results_ac31}(a) and \ref{results_ac32}(a)] 
the system has a gap between the upper and lower 
bands. When the ac field is switched on, the gap collapses [Fig.~\ref{results_ac31}(b)
and \ref{results_ac32}(b)], and a spectral weight grows in the midgap region 
around $\omega=0$. If we compare Figs.~\ref{results_ac31}(b) and 
\ref{results_ac32}(b), we can
see that the larger the band gap, the smaller the midgap weight. 
From these results, we see that metallic states appear in the 
insulating system of correlated electrons in the intense ac field. 
As the intensity of the ac field is further increased, 
the system plunges into the dynamical 
localization regime, where the band width starts to scale with $J_0(A)$. 

\begin{figure}[t]
  \begin{center}
    \includegraphics[width=8.5cm]{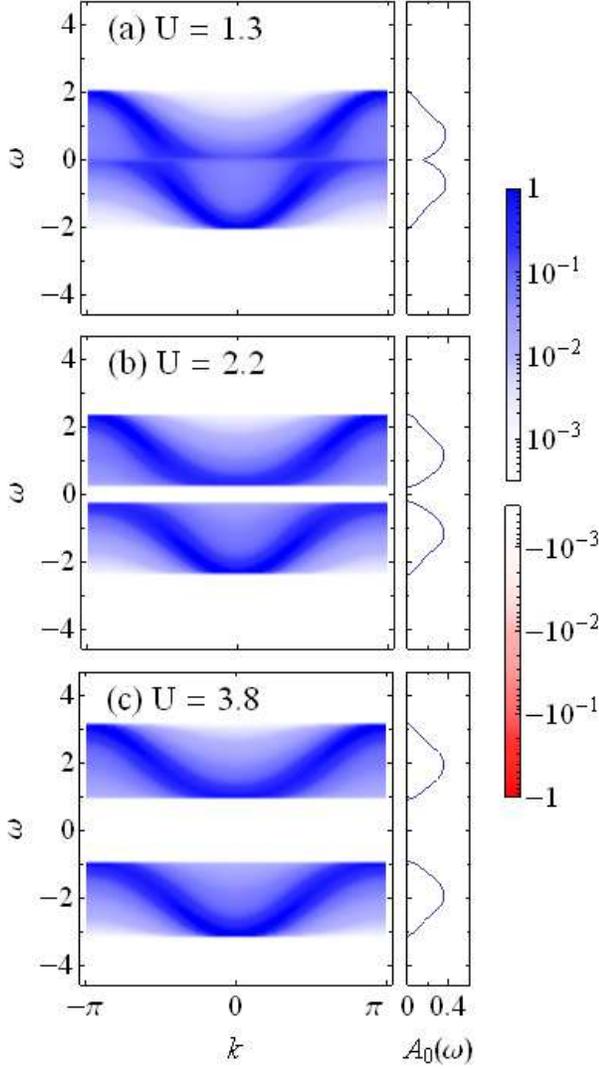}
    \caption{(Color online) The zeroth mode of the gauge-invariant spectral function 
		$\tilde{A}_0({\boldsymbol k}, \omega)$ with 
    ${\boldsymbol k} = k (1, 1, 1)$ (the density 
    plots) and the local spectral function $A_0(\omega)$ (the line plots)
    of the FK model coupled to the ac field ($\Omega=1$) on the 
    simple cubic lattice at half filling for $A=0$ in 
    units of $t^\ast$. The color bars on the right side
    represent the correspondence between the colors and the values of 
    $\tilde{A}_0({\boldsymbol k}, \omega)$. }
    \label{results_ac11}
  \end{center}
\end{figure}

\begin{figure}[t]
  \begin{center}
    \includegraphics[width=8.5cm]{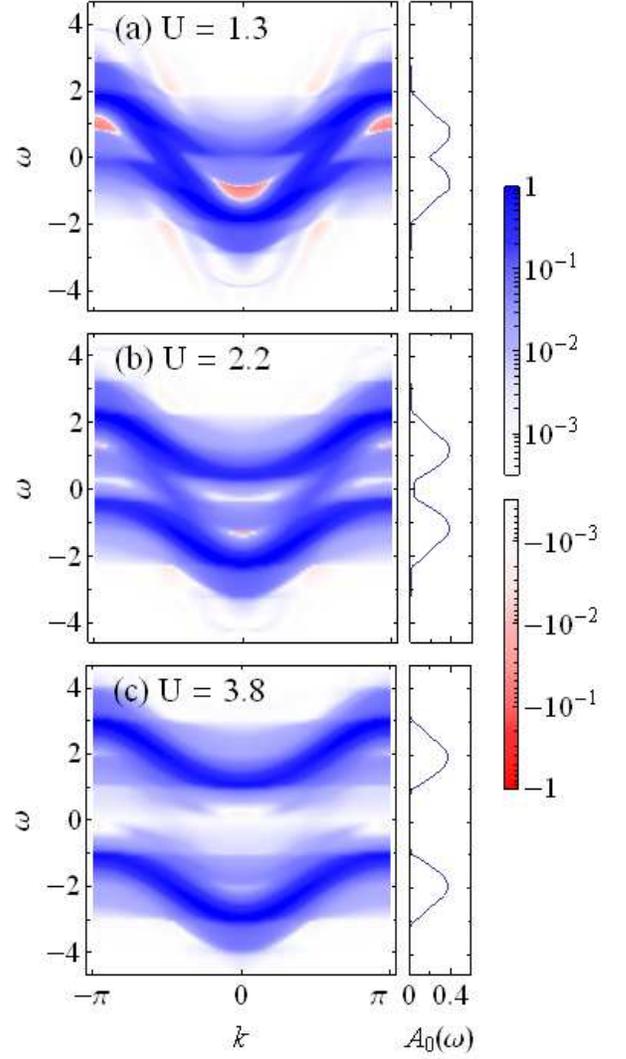}
    \caption{(Color online) The zeroth mode of the gauge-invariant spectral function 
		$\tilde{A}_0({\boldsymbol k}, \omega)$ with 
    ${\boldsymbol k} = k (1, 1, 1)$ (the density 
    plots) and the local spectral function $A_0(\omega)$ (the line plots)
    of the FK model coupled to the ac field ($\Omega=1$) on the 
    simple cubic lattice at half filling for $A=0.6$ in 
    units of $t^\ast$. The color bars on the right side
    represent the correspondence between the colors and the values of 
    $\tilde{A}_0({\boldsymbol k}, \omega)$. }
    \label{results_ac12}
  \end{center}
\end{figure}

To characterize the metallic state, we have calculated the 
momentum resolved spectral function $\tilde{A}_0({\boldsymbol k}, \omega)$. 
For clarity we take the simple cubic lattice 
with the JDOS [Eq.~(\ref{sc_jdos})] in three dimensions. 
As a key result in the present paper, 
we plot the zeroth mode of the spectral function $\tilde{A}_0({\boldsymbol k}, \omega)$ and 
$A_0(\omega)$ for the frequency $\Omega = 1$ at half filling in Figs.~\ref{results_ac11}
and \ref{results_ac12}, 
where we take ${\boldsymbol k}=k(1, 1, 1)$. One can check that the 
result respects the particle-hole symmetry. As discussed in 
Sec.~\ref{gauge_ac}, numerical evaluation of $\tilde{A}_0({\boldsymbol k}, \omega)$ 
is done in a perturbative way. Thus we can obtain reliable results only in a weak 
intensity region. Although we use a perturbation in $A$, higher-order contributions 
are included in our calculations. To obtain the results in Figs. 
\ref{results_ac11} and \ref{results_ac12}, the summation over $\ell$ in Eq.~(\ref{ac_gauge_flq}) is 
performed up to $\ell=5$. We have checked that the expansion in terms of 
$A$ converges for $A \lesssim 0.6$. 
A calculation tends to be unstable when $U$ is small 
($\lesssim 1$) where the system is in a metallic state. 
When we analyze the results, we have to be careful with the sign of the spectral function
$\tilde{A}_0({\boldsymbol k}, \omega)$. Although the local spectral function 
$A_0(\omega)$ is positive definite, $\tilde{A}_0({\boldsymbol k}, \omega)$ is 
not so in general. 
While we notice there are some regions where $\tilde{A}_0({\boldsymbol k}, \omega)$ 
becomes negative, the quantity is mostly positive 
for $A \lesssim 0.6$ and $U \gtrsim 2$. The result should be 
supported by other gauge-invariant quantities such as the current or the 
optical conductivity, which is a future problem. 

First, let us see Figs.~\ref{results_ac11}(a)-\ref{results_ac11}(c). These are the spectra 
of the system in equilibrium ($A=0$). One can see how the metallic bands 
[Fig.~\ref{results_ac11}(a)] change into the insulating ones 
[Figs.~\ref{results_ac11}(b) and \ref{results_ac11}(c)] with a finite gap appearing with $U$. 
In the insulating state, the upper band is almost a replica of the lower 
one shifted upward by $U$, 
which is characteristic of the FK model. 
When the ac field is turned on in Figs.~\ref{results_ac12}(a)-\ref{results_ac12}(c) ($A=0.6$), 
we can see how the ac field generates a new {\it photoinduced band structure}. 
In Fig.~\ref{results_ac12}(b), we can observe that a new 
band appears in the midgap region. This band is created by the electrons 
that absorb or emit one photon with the energy $\Omega=1$, 
that is, the photoinduced band is a replica of the original lower and 
upper bands shifted by $\Omega$. If we assume that the states are 
occupied up to $\omega=0$ as in equilibrium, 
the electrons in the induced band around $\omega=0$ play 
a role of carriers, making the system metallic. 
When the interaction $U$ is strong enough [Fig.~\ref{results_ac12}(c), $U=3.8$], 
the metallic band does not appear. Instead, 
side bands appear near the original bands with the spacing $\Omega$ in 
the midgap region. 
Again, the electrons in the side bands 
consist of the electrons absorbing or emitting one photon with the energy $\Omega$. 
Since $\Omega$ is much smaller than $U$ here, the electrons cannot reach the 
region around $\omega=0$ with a one-photon process. Therefore the system 
remains to be insulating with the finite gap in the ac field. 
As well as the case of the 
dc field, the side band pattern with the spacing 
$\Omega$ interferes with the original band structure with the spacing $U$, 
yielding complicated band patterns. 


\subsection{Relevance to experiments}

Finally we mention the relevance of the present results to 
experiments. For the dc field, the intensity required for the effect
considered here is $\sim 10^{9{\mathchar`-}10}$ V/m for $a\sim 10^{0{\mathchar`-}1}{\rm \AA}$, 
which is too strong to be realistic. However, in the case of the 
ac field, the required 
intensity is $A \sim 1$ [see Eq.~(\ref{ac_a})] for the dimensionless quantity, which 
translates to $E\sim 10^{9{\mathchar`-}10}$ V/m (i.e., the intensity of
$\sim 10^{11{\mathchar`-}13}$ W/cm$^2$) for $\Omega\sim$ 1 
eV (visible light). For a smaller $\Omega$ the required intensity becomes smaller. 
Since the intensity of pulsed laser available with recent advances in 
optical techniques reaches such magnitudes,\cite{mourou_tb2006} 
it should be possible to observe the nonlinear effects predicted in this paper 
in experiments. 
One problem is that when the intensity goes beyond $\sim 10^{12}$ W/cm$^2$, 
atoms begin to be ionized and evaporated. 
To make the required field intensity smaller, we can 
take systems with large lattice constants, such as 
the zeolites loaded with guest atoms.\cite{zeolite} 
As an entirely different class of systems, we can 
consider cold atoms in optical lattices,\cite{morsch_oberthaler2006} 
which may be an interesting playing ground for the effects examined in the present paper. 

\section{Conclusion}
\label{conclusion}

We have developed a theoretical method to formulate photoinduced 
phenomena in correlated electron systems. The method 
incorporates FGFM into DMFT, 
which can fully take into account both the electron correlation effect and 
the nonlinear electric-field effect. 
We have applied the method to the Falicov-Kimball model in ac fields to calculate 
the gauge-invariant spectral functions. 
We find peculiar photoinduced band structures, which arise from the 
nonlinear effect of the electric field. In particular, we find a 
metallic state in the midgap region of the 
Mott-like insulator induced by the ac field. 
In the calculation we have utilized 
a theorem, found here, that identifies eigenvalues and eigenvectors 
of a Floquet matrix for single-band noninteracting electrons. 

We believe that our approach has a potential ability to treat, not only the FK model 
considered in the paper, but also a wide class of models such as the Hubbard model. 
There are some future problems: one is 
to calculate the Keldysh component of Green's function $G^K$.
We need the Keldysh Green's function to compute, e.g., the current 
or the optical conductivity which has information of the transport 
properties of the system. An application to the Hubbard model
is also desirable. 
Experimentally, the relaxation of 
photoinduced states after the ac field is switched off 
is also an important phenomenon.   
This is theoretically interesting as well, for which 
further developments on the nonequilibrium DMFT would be necessary. 


\section*{Acknowledgment}
This work was supported in part by a Grant-in-Aid for Scientific Research 
on a Priority Area ``Anomalous quantum materials'' 
from the Japanese Ministry of Education.
N.T. was supported by the Japan Society for the Promotion of Science. 

\appendix
\section{Multiplication rule for the Floquet matrices}
\label{multiplication}

Here we show that Floquet matrices obey the multiplication rule 
in the linear algebra. Let us prepare two 
functions $A(t, t')$ and $B(t, t')$ which satisfy the 
periodicity condition: $A(t+\tau, t'+\tau)=A(t, t')$ (and so does $B$). We write the 
following integral in the Wigner representation: 
\begin{widetext}
\begin{align*}
      \int dt'' A(t, t'') B(t'', t') 
  &=
    \int dt'' \sum_\ell \int \frac{d\omega}{2\pi}
    e^{ 
      -i\omega(t-t'')-i\ell\Omega(t+t'')/2
    } 
    A_{\ell}(\omega)
    \sum_{\ell '}\int \frac{d\omega '}{2\pi}
    e^{
      -i\omega '(t''-t')-i\ell '\Omega(t''+t')/2
    }
    B_{\ell '}(\omega ')
  \\
  &=
    \sum_{\ell \ell '}
		\int \frac{d\omega}{2\pi} \int \frac{d\omega '}{2\pi}
    2\pi\delta\Big(\omega-\omega '-\frac{\ell+\ell '}{2}\Omega\Big)
    e^{
      -i\omega t-i\ell\Omega t/2+i\omega 't'-i\ell '\Omega t'/2
    }
    A_\ell(\omega)B_{\ell '}(\omega ')
	\\
  &=
    \sum_{\ell \ell '}\int \frac{d\omega}{2\pi}\;
    e^{
      -i\omega t-i\ell\Omega t/2
      +i[\omega-(\ell+\ell ')\Omega/2]t'-i\ell '\Omega t'/2
    }
    A_\ell(\omega)B_{\ell '}\Big(\omega-\frac{\ell+\ell '}{2}\Omega\Big)
  \\
  &=
    \sum_{\ell \ell '}\int \frac{d\omega}{2\pi}\;
    e^{
      -i(\omega-\ell '\Omega/2)(t-t')-i(\ell+\ell ')\Omega(t+t')/2
    }
    A_\ell(\omega)
    B_{\ell '}\Big(\omega-\frac{\ell+\ell '}{2}\Omega\Big). 
\end{align*}
\end{widetext}
Thus we have 
\begin{align*}
  (AB)_k(\omega)
	  =
		  \sum_{\ell+\ell '=k}
      A_\ell\Big(\omega+\frac{\ell '}{2}\Omega\Big)
      B_{\ell '}\Big(\omega-\frac{\ell}{2}\Omega\Big).
\end{align*}
Let us take an integer $k'$ satisfying the condition 
$k' \equiv k \;({\rm mod}\; 2)$. Replacing $\omega$ with 
$\omega+k'\Omega/2$ gives 
\begin{align}
  (AB)_k\Big(\omega+\frac{k'}{2}\Omega\Big)
    &=
      \sum_{\ell+\ell '=k}
      A_\ell\Big(\omega+\frac{k'+\ell '}{2}\Omega\Big)
	\nonumber
	\\
	  &\quad\times
      B_{\ell '}\Big(\omega+\frac{k'-\ell}{2}\Omega\Big).
	\label{wigner_ab}
\end{align}
If we write Eq.~(\ref{wigner_ab}) in the Floquet representation following 
its definition (\ref{floquet_rep}), we arrive at the conclusion, 
\begin{align*}
  &(AB)_{(k+k')/2, (k'-k)/2}(\omega)
	\\
	  &=
		  \sum_{\ell}
			A_{(k+k')/2, (k+k')/2-\ell}(\omega)
			B_{(k+k')/2-\ell, (k'-k)/2}(\omega),
	\nonumber
\end{align*}
or a more transparent expression, 
\begin{equation}
  (AB)_{mn}(\omega)
	  =
		  \sum_{m'} A_{mm'}(\omega) B_{m'n}(\omega), 
	\nonumber
\end{equation}
where $m = (k+k')/2$, $n = (k'-k)/2$, and $m' = (k+k')/2-\ell$,
all of which are integers due to $k\equiv k'\; ({\rm mod}\; 2)$. This 
ensures that we can apply the usual multiplication rule of a matrix to 
every Floquet-represented function.\\ 

\section{Derivation of the Floquet representation of the noninteracting Green's function}
\label{flq_g}

\begin{figure}[t]
  \begin{center}
	  \includegraphics[width=6cm]{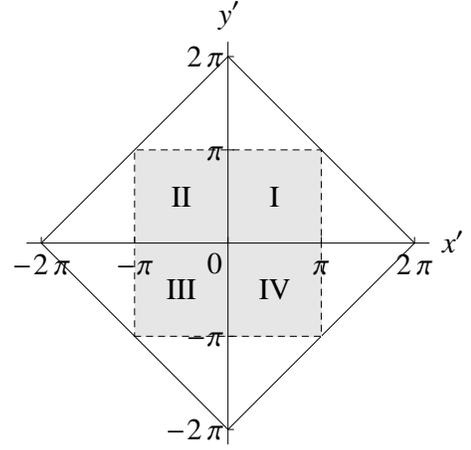}
    \caption{The range of the integral: each Roman numeral denotes the 
    corresponding shaded region, the symbols $\Diamond$ and $\Box$
    used in the text denote the regions surrounded 
		by the solid line and the broken line, respectively. }
		\label{range_of_integral}
		\end{center}
\end{figure}

Here we derive the Floquet representation 
[Eq.~(\ref{floquet1})] of Green's function. 
Let us start with Eq.~(\ref{retarded1}). 
We find that the 
argument of the exponential in Eq.~(\ref{retarded1}) is not invariant 
under discrete translation against $t_{\rm rel}$. To 
somehow make it invariant under such a translation, we rewrite Eq. 
(\ref{retarded1}) into 
\begin{align}
  G_{\boldsymbol k}^{R0}(t, t')
	  &=
		  -i\theta(t_{\rm rel}) e^{it_{\rm rel}[\mu-(\epsilon_{\boldsymbol k})_0]}
	\nonumber
	\\
	  &\times
			\exp\left(
			  -i \int_{t_{\rm av}-t_{\rm rel}/2}^{t_{\rm av}+t_{\rm rel}/2} dt'' \;
				[
				  \epsilon_{{\boldsymbol k}-e{\boldsymbol A}(t'' )} - (\epsilon_{\boldsymbol k})_0
				]
			\right).
	\label{retarded2}
\end{align}
Here $(\epsilon_{\boldsymbol k})_0$ is defined in Eq.~(\ref{e_flq}). 
Now that the argument of the exponential in Eq.~(\ref{retarded2})
is periodic in $t_{\rm rel}$ with the period $2\tau$ and in $t_{\rm av}$ 
with the period $\tau$, we can insert the factors 
$\sum_\ell e^{-i\ell\Omega t_{\rm rel}/2}\frac{1}{2\tau}\int_{-\tau}^{\tau}
dt_{\rm rel}'\;e^{i\ell\Omega t_{\rm rel}'/2}$ and 
$\sum_{m} e^{-im\Omega t_{\rm av}}\frac{1}{\tau}\int_{-\tau/2}^{\tau/2}
dt_{\rm av}'\;e^{im\Omega t_{\rm av}'}$ into Eq.~(\ref{retarded2}). 
Then, with the Fourier transformed expression of the step function, 
\begin{equation}
  \theta(t_{\rm rel})
	  =
		  -\frac{1}{2\pi i} \int d\omega ' \; 
			\frac{e^{-i\omega 't_{\rm rel}}}{\omega '+i\eta},
	\label{step}
\end{equation}
where $\eta$ is an infinitesimal positive constant, 
we can perform the Wigner transformation of Eq.~(\ref{retarded2}) as 
\begin{widetext}
\begin{align}
  (G_{\boldsymbol k}^{R0})_n(\omega)
	  =&
		  \sum_{\ell m} \int \frac{d\omega '}{2\pi} 
			\frac{1}{\omega '+i\eta} \int dt_{\rm rel} \;
			\frac{1}{\tau} \int_{-\tau/2}^{\tau/2} dt_{\rm av} \;
			e^{i[\omega+\mu-(\epsilon_{\boldsymbol k})_0-\omega '-\ell\Omega/2]t_{\rm rel}
			+i(n-m)\Omega t_{\rm av}}
	\nonumber
	\\
	  &\times
			\frac{1}{2\tau}\int_{-\tau}^{\tau} dt_{\rm rel}' \;
			\frac{1}{\tau} \int_{-\tau/2}^{\tau/2} dt_{\rm av}' \;
			e^{i\ell\Omega t_{\rm rel}'/2+im\Omega t_{\rm av}'}
			\exp\left(
			  -i\int_{t_{\rm av}'-t_{\rm rel}'/2}^{t_{\rm av}'+t_{\rm rel}'/2}
				dt'' \;
				[
				  \epsilon_{{\boldsymbol k}-e{\boldsymbol A}(t'' )}-(\epsilon_{\boldsymbol k})_0
				]
			\right). 
\end{align}
\end{widetext}
In order to make our notation clearer, 
we change the integral variables as $\Omega t_{\rm rel}'/2=x',\;\Omega t_{\rm av}'=y'$.
After some calculations, we obtain 
\begin{align}
  (G_{\boldsymbol k}^{R0})_n(\omega)
	  =&
		  \sum_{\substack{\ell\equiv n \\{\rm mod}\, 2}}
			\frac{1}{\omega-\ell\Omega/2+\mu-(\epsilon_{\boldsymbol k})_0+i\eta}
	\nonumber
	\\
	  &\times
			\int_{-\pi}^{\pi} \frac{dx'}{2\pi} \;
			\int_{-\pi}^{\pi} \frac{dy'}{2\pi} \;
			e^{i\ell x' +iny'}
	\nonumber
	\\
	  &\times
			  \exp\left(
				  -\frac{i}{\Omega} \int_{y'-x'}^{y'+x'} dz \;
					[
					  \epsilon_{{\boldsymbol k}-e{\boldsymbol A}(z/\Omega)} - (\epsilon_{\boldsymbol k})_0
					]
				\right). 
	\label{wigner1}
\end{align}
In the above we have used the fact that the integral in Eq.~(\ref{wigner1})
equals zero when $\ell\not\equiv n \;({\rm mod}\; 2)$ since 
$\iint dx' dy'=(\iint_{\rm I}+\iint_{\rm III})+(\iint_{\rm 
II}+\iint_{\rm IV})=[1+(-1)^{\ell+n}](\iint_{\rm I}+\iint_{\rm II})$, where the Roman 
numerals represent the ranges of the integral defined in Fig.~\ref{range_of_integral}.
We further change the integral variables: $x'+y'=x,\;x'-y'=-y$. Here we 
have to be careful with the range of the integral. As shown in Fig.~\ref{range_of_integral}, 
we change the range of the integral from $\iint_\Box dx'dy'$ to $\frac{1}{2}\iint_{\Diamond}dx'dy'$,
which is equal to $\frac{1}{2}\int_{-2\pi}^{2\pi}dx\int_{-2\pi}^{2\pi}dy$ times 
$\frac{1}{2}$ coming from the Jacobian. Then we change the 
range of the integral again: $\frac{1}{2}\times\frac{1}{2}\int_{-2\pi}^{2\pi}dx\int_{-2\pi}^{2\pi}dy
=\int_{-\pi}^{\pi}dx\int_{-\pi}^{\pi}dy$. Note that the two $\frac{1}{2}$ 
factors are canceled out by the change of the range of the integral. As a 
consequence, we arrive at the general Wigner representation of Green's 
function, 
\begin{align}
  (G_{\boldsymbol k}^{R0})_n(\omega)
	  &=
		  \sum_{\substack{\ell \equiv n \\{\rm mod}\, 2}}
			\frac{1}{\omega-\ell\Omega/2+\mu-(\epsilon_{\boldsymbol k})_0+i\eta}
	\nonumber
	\\
	  &\times
			\int_{-\pi}^{\pi} \frac{dx}{2\pi} 
			\int_{-\pi}^{\pi} \frac{dy}{2\pi} 
  			e^{i(\ell+n)x/2 - i(\ell-n)y/2}
	\nonumber
	\\
	  &\times
			  \exp\left(
				  -\frac{i}{\Omega} \int_{y}^{x} dz \;
					[
					  \epsilon_{{\boldsymbol k}-e{\boldsymbol A}(z/\Omega)} - (\epsilon_{\boldsymbol k})_0
					]
				\right).
\end{align}
Next, let us move on to the Floquet representation. Following the 
definition of the Floquet representation [Eq.~(\ref{floquet})], we have 
\begin{align}
  &(G_{\boldsymbol k}^{R0})_{mn}(\omega)
	\nonumber
	\\
	  &=
		  \sum_{\substack{\ell\equiv m-n\\{\rm mod}\, 2}}
			\frac{1}{\omega+(m+n-\ell)\Omega/2+\mu-(\epsilon_{\boldsymbol k})_0+i\eta}
	\nonumber
	\\
	  &\quad\times
			\int_{-\pi}^{\pi} \frac{dx}{2\pi} \int_{-\pi}^{\pi} \frac{dy}{2\pi}\;
			e^{i(\ell+m-n)x/2-i(\ell-m+n)y/2}
	\nonumber
	\\
	  &\quad\times
			\exp\left(
			  -\frac{i}{\Omega} \int_y^x dz \;
				[\epsilon_{{\boldsymbol k}-e{\boldsymbol A}(z/\Omega)}-(\epsilon_{\boldsymbol k})_0]
			\right).
	\nonumber
\end{align}
In the above we can replace $m+n-\ell$ with $2\ell$ due to $m+n-\ell 
\equiv m-n-\ell \equiv 0\; ({\rm mod}\; 2)$, which gives 
Eq.~(\ref{floquet1}).

\section{Unitarity of $\Lambda_{\boldsymbol k}$}
\label{unitarity}

We prove that $\Lambda_{\boldsymbol k}$ defined by Eq.~(\ref{lambda})
is a unitary matrix for any $\epsilon_{\boldsymbol k}$ and ${\boldsymbol A}(t)$ as 
\begin{align}
      \sum_\ell (\Lambda_{\boldsymbol k})_{m\ell} (\Lambda_{\boldsymbol k}^\dagger)_{\ell n}
	  &=
		  \int_{-\pi}^{\pi} \frac{dx}{2\pi} 
			\int_{-\pi}^{\pi} \frac{dy}{2\pi} \; \sum_\ell
			e^{i(mx-ny)-i\ell(x-y)}
	\nonumber
	\\
	  &\times
			\exp\left(
			  -\frac{i}{\Omega} \int_y^x dz \;
				[
				  \epsilon_{{\boldsymbol k}-e{\boldsymbol A}(z/\Omega)}-(\epsilon_{\boldsymbol k})_0
				]
			\right)
	\nonumber
	\\
	  &=
		  \int_{-\pi}^{\pi} \frac{dx}{2\pi} 
			\int_{-\pi}^{\pi} dy \; 
		  \delta(x-y) \; e^{i(mx-ny)}
	\nonumber
	\\
	  &\times
			\exp\left(
			  -\frac{i}{\Omega} \int_y^x dz \;
				[
				  \epsilon_{{\boldsymbol k}-e{\boldsymbol A}(z/\Omega)}-(\epsilon_{\boldsymbol k})_0
				]
			\right)
	\nonumber
	\\
	  &=
		  \int_{-\pi}^{\pi} \frac{dx}{2\pi} \; e^{i(m-n)x}
		=
		  \delta_{mn}. 
	\nonumber
\end{align}
Since $\Lambda_{\boldsymbol k}$ is nothing but a set of Floquet state 
vectors [Eq.~(\ref{theorem})], this unitarity relation indicates that the 
Floquet states $u_{\boldsymbol k}^{m-n}$ form an orthonormal and complete
basis. 

\section{Inverse of $G_{\boldsymbol k}^{R0}$}
\label{inverse_g}

The inverse of $(G_{\boldsymbol k}^{R0})_{mn}(\omega)$ is calculated 
for any $\epsilon_{\boldsymbol k}$
and ${\boldsymbol A}(t)$ as follows. First, using 
expression (\ref{lambda_q_lambda}) we have ${G_{\boldsymbol k}^{R0}}^{-1} = 
\Lambda_{\boldsymbol k} \cdot Q_{\boldsymbol k}^{-1}(\omega) \cdot 
\Lambda_{\boldsymbol k}^\dagger$ since $\Lambda_{\boldsymbol k}$ is unitary 
as proved in Appendix \ref{unitarity}. 
Among the terms in the diagonal matrix $Q_{\boldsymbol k}^{-1}(\omega)$,
$[\omega+\mu-(\epsilon_{\boldsymbol k})_0+i\eta] \delta_{mn}$ commutes with 
$\Lambda_{\boldsymbol k}$, giving a trivial result. The only nontrivial part, 
$n\Omega \,\delta_{mn}$, is evaluated as 
\begin{widetext}
\begin{align}
      \sum_\ell (\Lambda_{\boldsymbol k})_{m\ell} \; \ell\Omega \;
	    (\Lambda_{\boldsymbol k}^\dagger)_{\ell n}
	  =&
		  \int_{-\pi}^{\pi} \frac{dx}{2\pi} 
			\int_{-\pi}^{\pi} \frac{dy}{2\pi} \; \sum_\ell
			e^{i(mx-ny)-i\ell(x-y)} \; \ell\Omega \;
			\exp\left(
			  -\frac{i}{\Omega} \int_y^x dz \;
				[
				  \epsilon_{{\boldsymbol k}-e{\boldsymbol A}(z/\Omega)}-(\epsilon_{\boldsymbol k})_0
				]
			\right)
	\nonumber
	\\
	  =&
		  \int_{-\pi}^{\pi} \frac{dx}{2\pi} 
			\int_{-\pi}^{\pi} dy \; 
			i\Omega [\partial_x \delta(x-y)] e^{i(mx-ny)}
			\exp\left(
			  -\frac{i}{\Omega} \int_y^x dz \;
				[
				  \epsilon_{{\boldsymbol k}-e{\boldsymbol A}(z/\Omega)}-(\epsilon_{\boldsymbol k})_0
				]
			\right)
	\nonumber
	\\
	  =&
		  \int_{-\pi}^{\pi} \frac{dx}{2\pi} 
			\int_{-\pi}^{\pi} dy \; 
			\delta(x-y) 
			[m\Omega-\epsilon_{{\boldsymbol k}-e{\boldsymbol A}(x/\Omega)}+(\epsilon_{\boldsymbol k})_0]
      e^{i(mx-ny)} 
  \nonumber
	\\
	  &\times
			\exp\left(
			  -\frac{i}{\Omega} \int_y^x dz \;
				[
				  \epsilon_{{\boldsymbol k}-e{\boldsymbol A}(z/\Omega)}-(\epsilon_{\boldsymbol k})_0
				]
			\right)
	  =
		  [m\Omega+(\epsilon_{\boldsymbol k})_0]\delta_{mn} - (\epsilon_{\boldsymbol k})_{m-n}.
	\label{geometrical_phase}
\end{align}
\end{widetext}
Between the second and the third lines we have integrated by parts. The 
contribution coming from the boundary is negligible due to the presence 
of the delta function. Thus the 
simple expression [Eq.~(\ref{inv_g})] results. 

We note that relation (\ref{geometrical_phase}) for the Floquet 
indices $m=n=0$ reduces to the geometrical phase, 
$\gamma_{\boldsymbol k} = \int_{-\tau/2}^{\tau/2} dt\; u_{\boldsymbol k}^\ast(t)
i\partial_t u_{\boldsymbol k}(t)$. From Eq.~(\ref{geometrical_phase}), 
we can show that it is exactly absent in the single-band noninteracting 
system, $\gamma_{\boldsymbol k}= -2\pi \sum_\ell \ell\, 
(\Lambda_{\boldsymbol k})_{0\ell}(\Lambda_{\boldsymbol k}^\dagger)_{\ell 0} = 0$.

\section{Joint density of states in arbitrary finite dimensions}
\label{jdos}

\begin{figure}[t]
  \begin{center}
    \includegraphics[width=8cm]{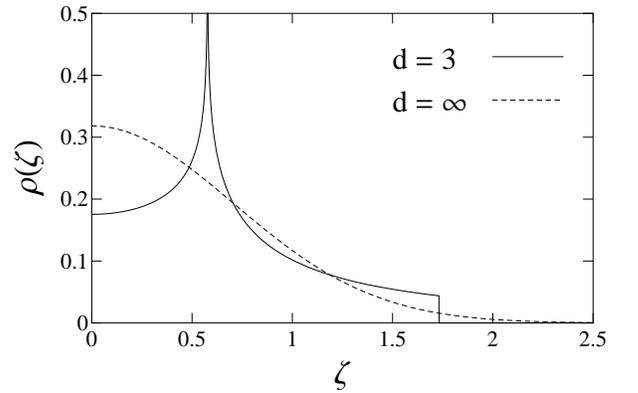}
		\caption{The JDOS $\rho(\zeta)$ for $d=3$ and $d=\infty$ in units of 
	 $t^\ast$. }
		\label{jdos_fig}
	\end{center}
\end{figure}
An analytic expression for JDOS defined by Eq.~(\ref{jdos_def}) can be 
obtained in arbitrary dimensions. We first substitute the delta 
functions in Eq.~(\ref{jdos_def}) with the integrals over auxiliary 
variables $s$ and $\bar{s}$: 
\begin{align*}
  \rho(\epsilon, \bar{\epsilon})
    &=
      \int_{-\infty}^\infty\frac{ds}{2\pi}\int_{-\infty}^\infty\frac{d\bar{s}}{2\pi}\; 
			e^{is\epsilon+i\bar{s}\bar{\epsilon}}
	\\
	  &\quad\times
      \sum_{\boldsymbol k}e^{2it\sum_i(s\cos k_i+\bar{s}\sin k_i)}.
\end{align*}
We then replace $\epsilon$ and $\bar{\epsilon}$ with $\zeta$ and 
$\theta$ in accordance with $\epsilon = \zeta \cos \theta$ and 
$\bar{\epsilon} = \zeta \sin \theta$, 
and change the integral variables as $s = \xi \sin \phi$ and 
$\bar{s} = \xi \cos \phi$.
After the integrations, we obtain 
\begin{equation}
  \rho(\zeta)
    =
      \int_0^\infty \frac{d\xi}{2\pi}\;\xi\;
      J_0(\zeta\xi)[J_0(2t\xi)]^d.
  \label{jdos_d}
\end{equation}
This is the general expression for the JDOS in $d$ dimensions. 
Note that the JDOS is independent of $\theta$ in any dimension. 
The infinite-dimensional case is readily reproduced since $J_0(z) = 1 - 
(z/2)^2 + O(z^4)$ and $t = t^\ast/2\sqrt{d}$, the factor $[J_0(2t\xi)]^d$ 
converges to $e^{-(\xi/2)^2}$ in the limit $d \to \infty$. 
With a formula for the integral of the Bessel 
function, we reproduce the known JDOS in infinite dimensions:\cite{turkowski_freericks2005}
\begin{equation}
  \rho_{d=\infty}(\zeta)
	  =
		  \frac{1}{\pi} e^{-\zeta^2}.
	\label{gaussian}
\end{equation}

In the case of finite dimensions, we can systematically deduce the 
JDOS from Eq.~(\ref{jdos_d}) with an appropriate integral formula 
related to the Bessel function. In the following we list the derived 
expression of the JDOS for $d=1, 2, 3$: 
\begin{align}
  \rho_{d=1}(\zeta)
    &=
      \frac{1}{2\pi(2t)}\delta(\zeta-2t),
  \\
  \rho_{d=2}(\zeta)
    &=
        \begin{cases}
          \dfrac{1}{\pi^2\zeta\sqrt{4(2t)^2-\zeta^2}} & 0 < \zeta < 4t
          \\
          0 & 4t < \zeta
        \end{cases},
\end{align}
\begin{widetext}
\begin{align}
  \rho_{d=3}(\zeta)
    &=
      \begin{cases}
			    \dfrac{2}{\pi^3\sqrt{(\zeta+2t)^3(6t-\zeta)}}\;K\!\left(
          4\sqrt{\dfrac{(2t)^3\zeta}{(\zeta+2t)^3(6t-\zeta)}}\right) 
        & 
          0 \leq \zeta < 2t
        \vspace{.3cm}
        \\
          \dfrac{1}{2\pi^3\sqrt{(2t)^3\zeta}}\;K\!\left(
          \dfrac{1}{4}\sqrt{\dfrac{(\zeta+2t)^3(6t-\zeta)}{(2t)^3\zeta}}\right)
        &
          2t < \zeta \leq 6t
        \vspace{.3cm}
        \\
          0 
        &
          6t < \zeta
      \end{cases},
	\label{sc_jdos}
\end{align}
where $K(k)$ is the elliptic integral of the first kind. In Fig.~\ref{jdos_fig} 
we plot the JDOS for $d=3$ and $d=\infty$. We can 
observe that $\rho_{d=3}$ diverges at $\zeta = 2t$, which originates 
from the van Hove singularity of the simple cubic lattice. 
\end{widetext}

\bibliographystyle{apsrev}
\bibliography{floquet}

\end{document}